\documentclass{jfm}
\usepackage{color}
\usepackage{graphicx}
\usepackage{natbib}
\usepackage{mathtools}

\ifCUPmtlplainloaded \else
  \checkfont{eurm10}
  \iffontfound
    \IfFileExists{upmath.sty}
      {\typeout{^^JFound AMS Euler Roman fonts on the system,
                   using the 'upmath' package.^^J}%
       \usepackage{upmath}}
      {\typeout{^^JFound AMS Euler Roman fonts on the system, but you
                   dont seem to have the}%
       \typeout{'upmath' package installed. JFM.cls can take advantage
                 of these fonts,^^Jif you use 'upmath' package.^^J}%
      }
  \else
  \fi
\fi

\ifCUPmtlplainloaded \else
  \checkfont{msam10}
  \iffontfound
    \IfFileExists{amssymb.sty}
      {\typeout{^^JFound AMS Symbol fonts on the system, using the
                'amssymb' package.^^J}%
       \usepackage{amssymb}%
       \let\le=\leqslant  
         
      }{}
  \fi
\fi

\ifCUPmtlplainloaded \else
  \IfFileExists{amsbsy.sty}
    {\typeout{^^JFound the 'amsbsy' package on the system, using it.^^J}%
     \usepackage{amsbsy}}
    {\providecommand\boldsymbol[1]{\mbox{\boldmath $##1$}}}
\fi

\newsavebox{\astrutbox}
\sbox{\astrutbox}{\rule[-5pt]{0pt}{20pt}}

\title[DNS of TC flow with grooved walls]{Direct numerical simulation of Taylor-Couette flow with grooved walls: torque scaling and flow structure}

\author[X. Zhu, R. Ostilla-M$\text {\'{o}}$nico, R. Verzicco and D. Lohse]%
{Xiaojue Zhu$^1$%
  ,\ns
Rodolfo Ostilla-M$\text {\'{o}}$nico$^1$%
 ,\ns
Roberto Verzicco$^{1,2}$%
 ,\ns
and Detlef Lohse$^1$}

\affiliation{$^1$Physics of Fluids Group, MESA+ Institute and J. M. Burgers Centre for Fluid Dynamics, University of Twente, P.O. Box 217, 7500AE Enschede, The Netherlands\\[\affilskip]
$^2$Dipartimento di Ingegneria Industriale, University of Rome "Tor Vergata",
Via del Politecnico 1, Roma 00133, Italy}

\pubyear{2010}
\volume{650}
\pagerange{119--126}
\date{?; revised ?; accepted ?. - To be entered by editorial office}
\begin{document}

\maketitle

\begin{abstract} 
We present direct numerical simulations of Taylor-Couette flow with grooved walls at a fixed radius ratio $\eta=r_i/r_o=0.714$ with inner cylinder Reynolds number up to $Re_i=3.76\times10^4$, corresponding to Taylor number up to $Ta=2.15\times10^9$. The grooves are axisymmetric V-shaped obstacles attached to the wall with a tip angle of $90^\circ$. Results are compared to the smooth wall case in order to investigate the effects of grooves on Taylor-Couette flow. We focus on the effective scaling laws for the torque, flow structures, and boundary layers. It is found that, when the groove height is smaller than the boundary layer thickness, the torque is the same as that of the smooth wall cases. With increasing $Ta$, the boundary layer thickness becomes smaller than the groove height. Plumes are ejected from the tips of the grooves and secondary circulations between the latter are formed. This is associated to a sharp increase of the torque and thus the effective scaling law for the torque vs. $Ta$ becomes much steeper. Further increasing $Ta$ does not result in an additional slope increase. Instead, the effective scaling law saturates to the ``ultimate'' regime effective exponents seen for smooth walls. It is found that even though after saturation the slope is the same as for the smooth wall case, the absolute value of torque is increased, and the more the larger size of the grooves.
\end{abstract}


\section{Introduction}
\label{sec1}

Non-smooth surfaces exist everywhere in nature, and many engineering applications need to deal with rough boundaries. The question of how local wall roughness affects global transport properties dates back to the pioneering study by \cite{nikuradse1933} in pipe flow. Nikuradse performed experiments on pipes with sand glued to the wall as densely as possible. The measurements of the friction coefficient $C_f=\tau/(\frac{1}{2}\rho U^2)$, where $\tau$ is the surfaced averaged friction stress, $\rho$ the fluid density and $U$ the mean flow velocity, shows that roughness has little impact in the laminar regime but after increasing the Reynolds number $Re$, the friction factor turns upward and reaches an asymptote. At the highest $Re$, the friction factor becomes independent of $Re$. Nikuradse then explained that in the smooth case, the viscous sublayer depth depends on $Re$, and hence the friction factor. However, by introducing roughness the viscous sublayer decreases down to the roughness scale, where the friction factor becomes independent of the $Re$. Since then, there have been a lot of studies concerning flow in pipes with roughed surfaces (see \cite{jimenez2004} for a review).

In general, studying the effect of a change of the boundary conditions at the wall will lead to a better understanding of the bulk-boundary layer (BL) interaction and the flow transport properties which are closely connected therewith. Next to pipe flow, the canonical systems in turbulent flows are Rayleigh-B$\text {\'{e}}$nard (RB) flow, in which a fluid is driven by the temperature difference between the hot bottom plate and cold top plate, and Taylor-Couette (TC) flow, in which a fluid is confined between two independently rotating coaxial cylinders. Both flows have been well studied and show rich patterns already with smooth walls (see \cite*{ahlers2009} for a comprehensive review on RB and \cite{fardin2014}; \cite*{grossmann2016} on TC). \citet*{eckhardt2007a,eckhardt2007b} showed that pipe, RB and TC flow are analogous to each other. Because of the close analogy, a better understanding of TC will lead to a more profound insight also in RB and pipe flow and vice versa.

The temperature difference between the top and bottom plate in RB flow is analogous to different rotation rates of the inner and outer cylinders in TC flow. The rotation difference in TC flow is non-dimensionally characterised by the Taylor number $Ta$, which is analogous to the dimensionless temperature difference in RB, i.e. the Rayleigh number $Ra$ . For TC flow, the global transport property is expressed as dimensionless torque $Nu_{\omega}$, which is analogous to the dimensionless heat flux in RB, i.e. the Nusselt number $Nu$. In TC flow, when the driving force $Ta$ is small, both the BL and the bulk are of laminar type. When increasing $Ta$, first the bulk becomes turbulent and finally also the BLs \citep{grossmann2014,ostilla2014a,ostilla2014b}. This state is the so-called ``ultimate'' regime. The ultimate regime is relevant not only conceptually \citep{kraichnan1962,grossmann2000,grossmann2001} but also as many applications in nature and engineering are within that regime. For TC flow, the ``ultimate'' regime was first found by \cite{lathrop1992a,lathrop1992b,lewis1999}, though they did not call it like this. Later, \cite{vangils2011,huisman2012,huisman2013,ostilla2014a,ostilla2014b} put it into this conceptual framework. In RB turbulence, the ultimate regime was experimentally found by \cite{he2012a,he2012b}. In both RB and TC turbulence, the ultimate regime scalings, namely $Nu\sim Ra^{\beta}$ in RB flow and $Nu_{\omega} \sim Ta^{\beta}$ in TC flow, have an effective exponent around $\beta \approx 0.38-0.40$, originating from 1/2 \citep{kraichnan1962} and logarithmic corrections \citep{grossmann2011,grossmann2012}.

For RB flow with roughness, various different effective scaling laws relating heat transport to the driving, written in the form $Nu=ARa^{\beta}$, were suggested. When the height of roughness $\delta$ is larger than the thermal BL thickness $\lambda_{\theta}  \simeq L/(2Nu)$, where $L$ the distance between two plates, \cite{shen1996} found that the prefactor $A$ was increased by 20\% whereas the exponent $\beta$ did not change by using rough surfaces made of a regularly spaced pyramids. Later, by using the same facility but different pyramid height (9 mm compared to 3.2 mm in \citet{shen1996}), \cite{du2000} measured the increase of $A$ to be as much as 76\% and the exponent $\beta$ again stayed the same. Based on flow visualisation, \cite{du2000} concluded that the enhancement of heat transport is due to the plume ejection from the tip of the pyramids. Also \cite{ciliberto1999} found that $\beta$ was unaffected but more surprisingly $A$ was even decreased when $\lambda_{\theta} < \delta$. In another experiment, which used pyramid roughness, by \cite{qiu2005}, both $A$ and $\beta$ were found to increase and the new $\beta$ with roughness was 0.35. \cite{wei2014} found $\beta \approx 0.35$ with roughness on both the lower and upper plates. In contrast, \cite{stringano2006}, who numerically investigated RB convection over grooved plates, showed that the secondary vortex inside the grooves would lift up the BL and help the plumes detach from the tip, which is consistent with the result of \cite{du2000}. Both $A$ and $\beta$ were increased and $\beta$ was changed to be around 0.37. By implementing V-shaped axis-symmetrical roughness both on the side walls and horizontal plates, \cite{roche2001} obtained an increase of $\beta$ to be around 0.51, which was interpreted as triggering the ultimate region $1/2$ law without the logarithmic correction proposed by \cite{kraichnan1962}, after $\lambda_{\theta}$ is below the roughness height. They concluded that the roughness imposes a new length scale to the thermal BLs. They argued that the sublayer thickness would be fixed by the roughness so that logarithmic correction would become irrelevant. \cite*{ahlers2009} pointed out the 1/2 scaling observed might possibly be due to a crossover between rough surfaces with a groove depth less than the BL thickness to a regime where the groove depth is larger than the BL thickness. \cite{tisserand2011} postulated that if this interpretation was correct, then $\beta=1/2$ behaviour of RB with roughness would be fortuitous. \cite{salort2014} further showed that in their model $\beta$ could range from $0.38 \sim 0.5$ depending on the extent of instability of the BL. Clearly, more work is needed to resolve this issue.

We now come to TC flow with roughness, the subject of the present study, in which the studies are more rare. \cite{cadot1997} peformed experiments with equidistance ribs on both the inner and the outer surface. These ribs were straight and parallel to the axis of the cylinders. With smooth boundaries, the dissipation in the boundary is dominant and then drag coefficient decreases with increasing $Re$. However, with rough walls, \cite{cadot1997} argued that the dissipation in the BLs is no longer dominant, due to the extra dissipation in the bulk. In that regime the global drag coefficient becomes constant with increasing $Re$. Inspired by this work, \cite{berg2003} performed further experiments with the same style of roughness. They reported results for the four cases of two smooth walls, smooth-inner/rough-outer, rough-outer/smooth-inner, and two rough walls. The data were interpreted within the Grossmann-Lohse theory \citep{grossmann2000,grossmann2001,grossmann2002}. The flow was found to change from BL dominant to bulk dominant. In the case with two rough walls, the drag coefficient is again found to be independent of $Re$.
If translating $Re$ to $Ta$ and the drag coefficient to $Nu_{\omega}$, the effective scaling exponent $\beta$ in the relation $Nu_{\omega} \sim Ta^{\beta}$ is nearly 1/2 in the rough-rough case. The phenomenon of drag saturation with increasing $Re$ is very similar to what \cite{nikuradse1933} had found in his rough pipe experiments.

We stress that there are usually two different types of wall roughness. One is to arrange the roughness in a way to impede the mean flow. We call this ``perpendicular roughness''. This kind of roughness elements seem to be more efficient generators of skin friction than smooth walls \citep{jimenez2004}. The studies of \cite{cadot1997} and \cite{berg2003} can both be included in this category. The other possibility is to arrange the roughness in the way aligned to the mean flow, i.e. ``parallel roughness''. A well-documented example is the flow over riblets \citep{choi1993,chu1993}. Under specific circumstances they decrease drag by 6\% \citep{choi1993}. Many different kinds of roughness can be formed by combining these two ways.

\begin{figure}
\centering
\begin{minipage}[c]{0.46\textwidth}
\vspace{0.1cm}
\includegraphics[width=2.5in]{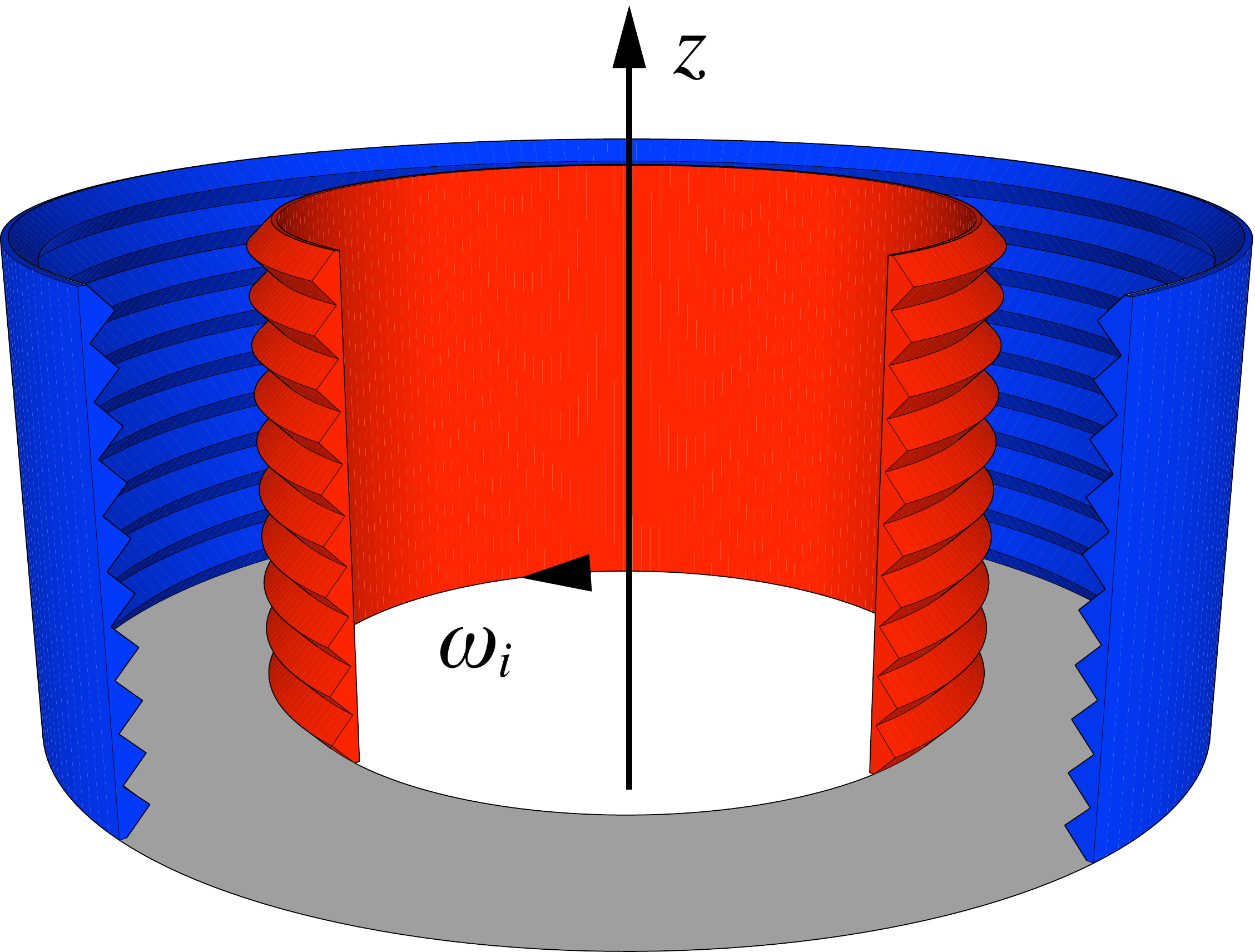}
\vspace{0.3cm}
    \centering {\par (a) \par}
\end{minipage}    
\begin{minipage}[c]{0.46\textwidth}
\vspace{0.cm}
\includegraphics[width=2.0in]{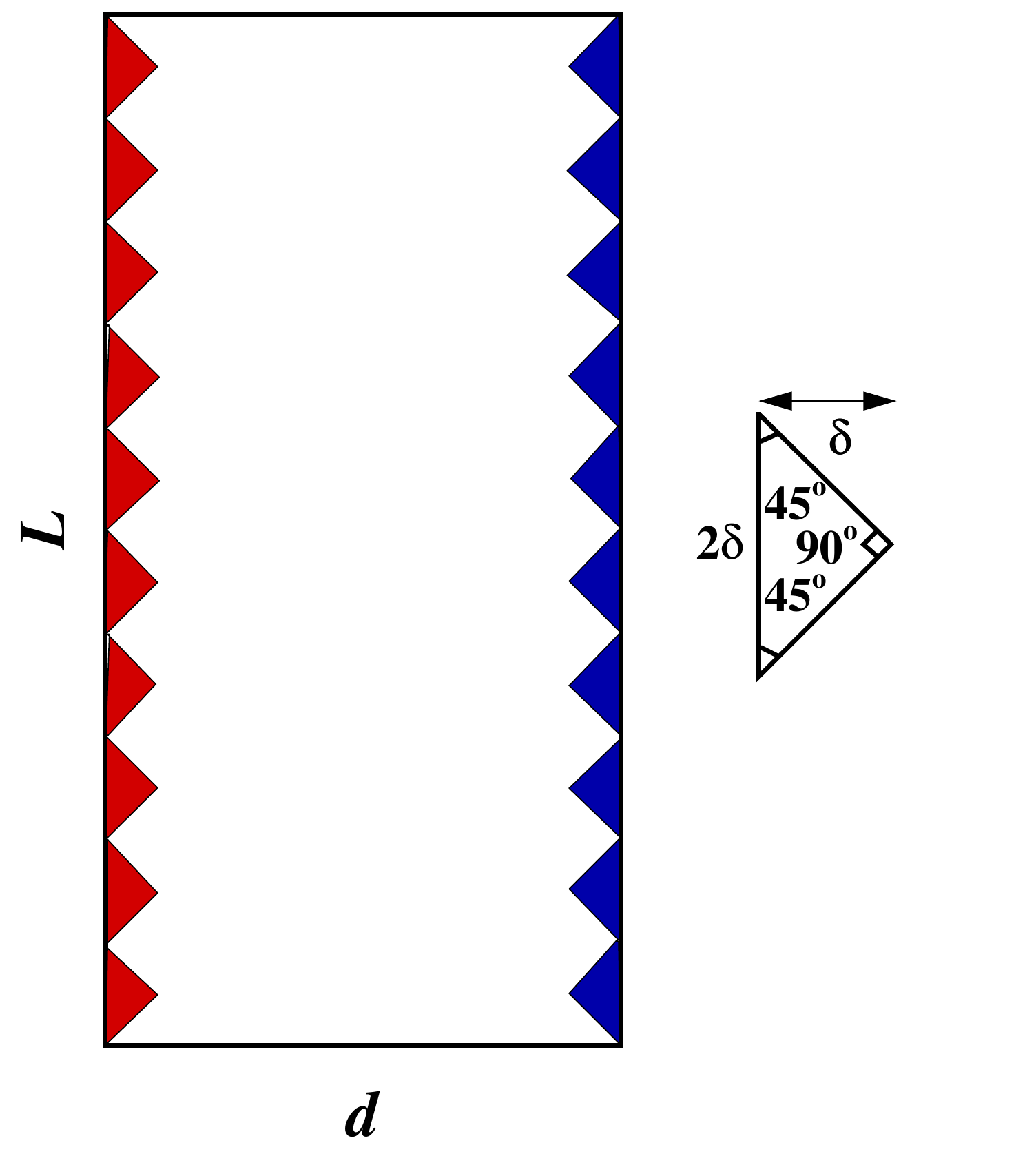}
\vspace{-0.15cm}
    \centering {\par (b) \par}
\end{minipage}   
 \caption{Schematic view of the Taylor-Couette system and the groove geometry. (a) Three dimensional view. The inner cylinder with radius $r_i$ is rotating with angular velocity $\omega_i$. The outer cylinder with radius $r_o$ is at rest. (b) Cross-section view of the the gap between the cylinders: $d=r_o-r_i$. The length scale of the grooves is $\delta$.}
\label{fig1}
\end{figure}

Inspired by the above studies, especially the similarities between pipe, RB, and TC flow, we study how the $Nu_\omega$ versus $Ta$ scaling and the corresponding flow structure behave with roughness in TC.  In the present study, we perform direct numerical simulation (DNS) of the TC flow with grooved walls. Because the grooves are quite large compared to the viscous scale, we avoid calling them roughness similar to \cite{stringano2006}. The grooves implemented here are V-shaped obstacles attached to the wall with a tip angle of $90^\circ$ and axisymetric to the axis of cylinder. This arrangement of grooves is categorised as parallel roughness. A schematic view of the structure is shown in figure \ref{fig1}. Note that the rough element type in this study was similar to the one used in \cite{stringano2006}, but different from the one of \cite{shen1996,du2000}, in which pyramids structures were used.

Our motivations are as follows: (i) DNS provides us with the ability to reproduce all the details of the flow field which are unavailable in experiments, and therefore enables us to reveal the connection between the effective scaling laws for the torque, the boundary layer, and the flow structures. (ii) We want to answer the question whether with parallel roughness the ultimate regime effective scaling exponent $\beta$ becomes 1/2 or whether it stays the same as in the smooth case, namely $\beta=0.38$ in the relevant $Ta$ regime, but with a different roughness strength dependent prefactor.  

The manuscript is organised as follows. In \S \ref{sec2}, we describe the numerical methods and parameter settings. In \S \ref{sec3}, we show the effective scaling laws between the Nusselt number and Taylor number in smooth and grooved cases. In \S \ref{sec4}, it is shown how grooves change the flow structure. \S \ref{sec5} presents the boundary layer dynamics with grooves. Finally, conclusions are drawn in  \S \ref{sec6}.
 
\section{Numerical settings}
\label{sec2}
\subsection{Parameter descriptions}

In the present study, the outer cylinder is stationary and only the inner cylinder is rotating and thus driving the flow. The flow is bounded by two lateral grooved walls cylinders with no-slip boundary conditions. The lower and upper surfaces are replaced by axially periodic boundary conditions and therefore do not include the effects of end walls unavoidably present in TC experiments. The grooves of the lateral walls are V-shaped with tip angle of $90^\circ$ and height $\delta$ (Figure \ref{fig1}). $r_i$ and $r_o$ are the base radii of the inner and the outer cylinder without grooves, respectively. The valley-to-valley distance $d=r_o - r_i$ is used to non-dimensionalize all lengths and the base velocity of the inner cylinder $U=r_i \omega_i$ for normalizing velocities, where $\omega_i$ is the angular velocity of the inner cylinder. The inner grooves rotate with the inner cylinder, and thus have constant angular velocity. This means that the \emph{azimuthal} velocity at the tip of the groove is slightly larger by factor of $(r_i+\delta)/r_i$ than the velocity at the valley of the inner cylinder. The geometry of the system is fixed at a specific radius ratio $\eta=r_i/r_o=0.714$. The reason for keeping the outer cylinder stationary and choosing such radius ratio are because they are close to the previous experimental and numerical studies \citep{lathrop1992a,lathrop1992b,lewis1999,ostilla2014a,huisman2012,huisman2013} so that we can make direct comparisons with those results. We define the dimensionless radial coordinate as $y=(r-r_i)/d$ so that it ranges from 0 at the inner cylinder to 1 at the outer cylinder. \cite{braukmann2013} and \cite{ostilla2014a,ostilla2014b} showed that a rotational symmetry of order 6 does not change the flow statistics for $\eta=0.714$. We follow their approach and choose this value to reduce the grids and thus computational cost. The aspect ratio $\Gamma$ is chosen to be $\Gamma=L/d=2\pi/3=2.094$ \citep{ostilla2014a}, where $L$ is the axial domain length. In this way one pair of Taylor vortices can be sustained in our DNS. The dimensionless torque is defined in the form of $Nu_\omega=T/T_{pa}$, where $T_{pa}$ is the torque of the purely azimuthal laminar state without grooves.

The motion of the fluid is governed by the incompressible Navier-Stokes equations
\begin{eqnarray}
\label{equ1}
\frac{\partial \bf{u}}{\partial t}+{\bf{u}} \cdot \nabla {\bf u} &=&- \nabla p + \frac{f(\eta)}{Ta^{1/2}}{\nabla}^2 {\bf u}, \\
\nabla  \cdot {\bf u} &=& 0, 
\end{eqnarray}
where $\bf u$ and $p$ are the fluid velocity and pressure, respectively. $f(\eta)$ is a geometrical factor which is in the form 
\begin{eqnarray}
f(\eta)= \frac{(1+\eta)^3}{8\eta^2}.
\end{eqnarray}
The $Ta$ number, with the absence of outer cylinder rotation, is written as 
\begin{equation}
Ta=\frac{1}{64} \frac{(1+\eta)^4}{\eta^2}d^2{(r_i+r_o)}^2\omega_i^2{\nu}^{-2},
\end{equation}
where $\nu$ is the kinematic viscosity of the fluid.

Another alternative way to determine the system by using the inner cylinder Reynolds number $Re_i=r_i \omega_i d/ \nu$, not by $Ta$ number, is suggested in the work of \citet{lathrop1992a,lathrop1992b,lewis1999}. Note that these two definitions can be easily transferred between each other by the relation
\begin{eqnarray}
Ta=[f(\eta)Re_i]^2.
\end{eqnarray}

\subsection{Numerical method}

A second-order finite difference code is employed for the present research, which is written in cylindrical coordinates and discretized on a staggered mesh. Details of the base code can be found in \cite{verzicco1996,vandepoel2014}. The code has been extensively validated in \cite{ostilla2013,ostilla2014a,ostilla2014b}. Time marching is performed by a third-order Runge-Kutta scheme and the fractional step is used for the pressure-momentum coupling, in combination with a semi-implicit scheme for viscous terms. To achieve large scale computation, a hybrid MPI-pencil and OpenMP decomposition is used to parallelize the code.

An immersed boundary (IB) technique \citep{fadlun2000} has been implemented into the code in order to deal with grooves on the surfaces of both cylinders. The main idea of the IB method is to add a body force term $\bf f$ to the momentum equation (\ref{equ1}), mimicking the boundary effect, to enforce in this way the desired velocity on the boundary, so that a regular non-body fitted mesh can be used. The information transfer between boundaries and nearby meshes is performed by means of interpolation. The advantage of IB is immediate: Flow bounded by arbitrary complex geometry can be easily solved on a very simple mesh with an additional body force. This IB method has already been validated through varieties of contexts \citep{fadlun2000,stringano2006}. For more details on the implementation, accuracy, and application of the IB method, we refer the reader to \cite{fadlun2000} and \cite{mittal2005}. 

\begin{figure}
  \centering
  \begin{minipage}[c]{0.48\textwidth}
    \includegraphics[width=1.2in]{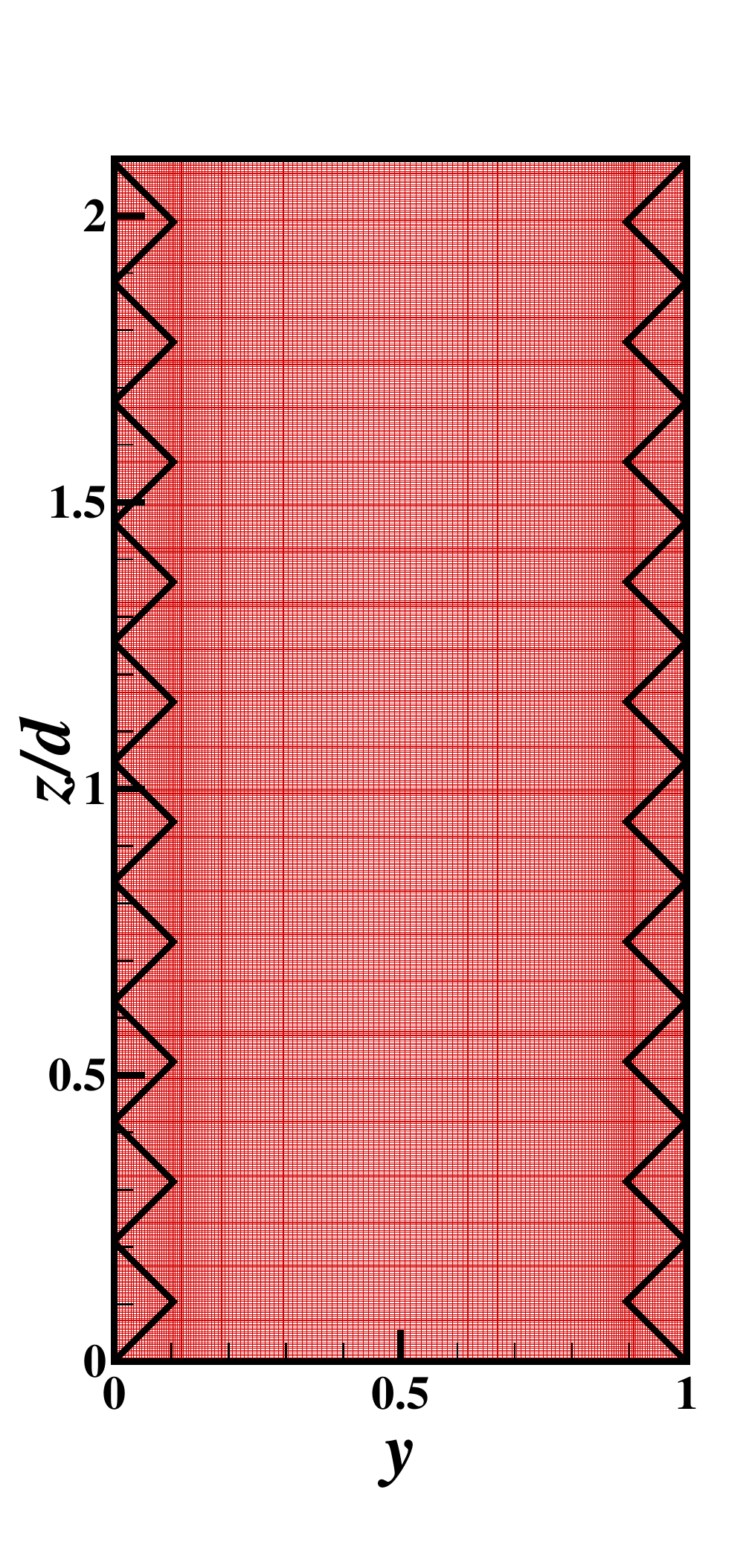}
    \centering {\par (a) \par}
  \end{minipage}%
     \begin{minipage}[c]{0.48\textwidth}
    \includegraphics[width=1.2in]{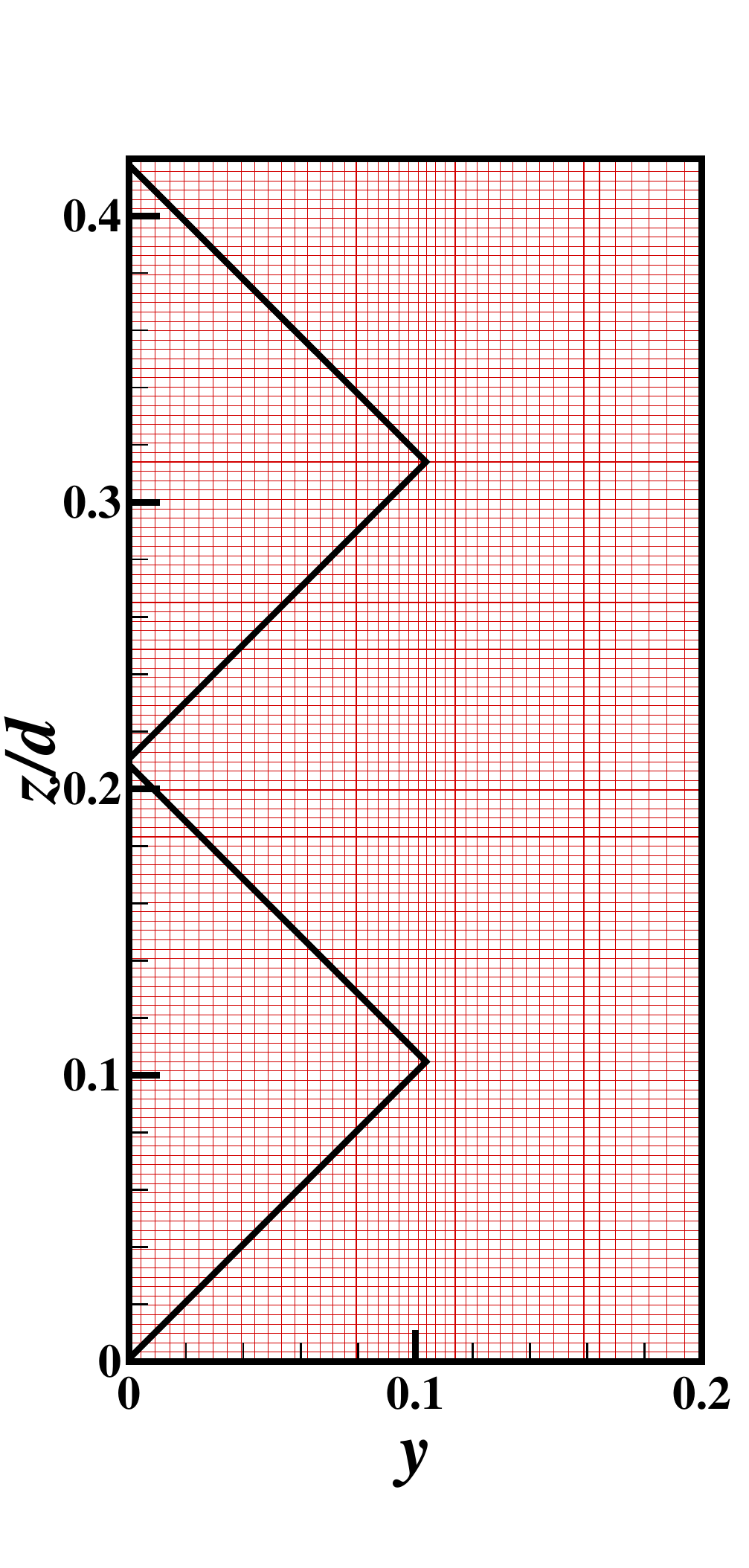}
    \centering {\par (b) \par}
  \end{minipage}
    \caption{An example of the mesh used in our simulations at meridional plane with radial grids and axial grids of $N_r \times N_z=160\times512$. (a) Whole surface. (b) Enlargement of the region near the grooves. A more refined mesh is needed near the tip of the grooves. Note that the grid resolution in axial direction is much higher than the one used in \cite{ostilla2013,ostilla2014a} because here we have to implement the IB method.}
\label{fig2}
\end{figure}

In order to guarantee the proper resolution, we proceed as follows. The mean angular velocity current $J^\omega$, defined by
\begin{eqnarray}
J^\omega=r^3(\langle u_r\omega \rangle_{A,t} - \nu \partial_r \langle \omega \rangle_{A,t}),
\end{eqnarray}
is strictly conserved along the radius $r$ \citep*{eckhardt2007b}. In this equation, $u_r$ denotes the radial velocity component and $\omega$ the angular velocity. On the one hand, $J^\omega$ is related to the torque $T$ by $T/(2\pi L \rho_f)=J^\omega$, where $\rho_f$ is the fluid density. On the other hand, $J^{\omega}$ can also be related to $Nu_{\omega}$ as $Nu_{\omega}=T/T_{pa}=J^{\omega}/J_0^{\omega}$, in which $J_0^{\omega}$ is the angular velocity current of the purely azimuthal state without grooves. $\langle ... \rangle_{A,t}$ represents averages over a cylindrical surface at radius $r$ and over time. Numerically, $J^{\omega}$ will deviate slightly from constant due to numerical errors. To quantify this difference, we define
\begin{eqnarray}
\Delta _J = \frac{\max(J^{\omega}(r))-\min(J^{\omega}(r))}{\langle J^{\omega}(r) \rangle_r},
\end{eqnarray}
where the maximum and minimum are determined over all $r$, which is chosen to be within the range of $r_i+\delta \le r \le r_o-\delta$ because of the influences of the grooves. As illustrated by \cite{ostilla2013}, $\Delta _J \le 0.01$ is a very strict requirement for the meshes. We make sure that all of our DNSs meet this criterion (see table \ref{tab1}). An additional issue on resolution within the near wall region by using IB method is that grooves do not coincide with the coordinate lines. This results in a finer mesh in the radial and axial directions. Compared to the case without grooves at the same $Ta$ number, the number of grids increases at least by a factor of 4. Figure \ref{fig2} shows an example of the mesh in a meridional plane.
 
 \begin{table}
\begin{center}
  \begin{tabular}{cccccc}
   $\delta/d$ & $Ta$ & $Re_i$ & $N_\theta \times N_r \times N_z$ & $Nu_\omega$ & $100\Delta _J $  \\ [3pt]
   
0.052& $2.44\times10^5$ & $4.00\times10^2$ & $64\times160\times512$ & 2.869  & 0.21   \\
0.052& $7.04\times10^5 $& $6.80\times10^2$ & $64\times160\times512$ &  3.655 & 0.42  \\
           0.052& $1.91\times10^6 $ & $1.12\times10^3$& $64\times192\times768$ & 4.678  &  0.33 \\
           0.052& $3.90\times10^6 $ & $1.60\times10^3$ & $128\times224\times896$ &  5.535 &  0.26 \\
           0.052& $9.52\times10^6 $ & $2.50\times10^3$ & $128\times256\times1024$ & 6.646  &  0.41 \\
           0.052& $2.39\times10^7 $ & $3.96\times10^3$ &$160\times320\times1280$ & 7.208  &  0.38 \\
           0.052& $4.77\times10^7 $ & $5.60\times10^3$ & $160\times400\times1600$ &  8.421 & 0.62\\
           0.052& $9.75\times10^7 $ & $8.00\times10^3$ & $192\times512\times2048$ &  11.11 & 0.56 \\
            0.052& $2.15\times10^8 $ & $1.19\times10^4$ &$192\times512\times2048$ &  16.60 & 0.82 \\
            0.052& $4.62\times10^8 $ & $1.74\times10^4$ & $256\times640\times2560$ & 22.70 & 0.79\\
            0.052& $9.75\times10^8 $ & $2.53\times10^4$ & $256\times640\times2560$ & 30.85  & 0.83 \\
            0.052& $2.15\times10^9 $ & $3.76\times10^4$ & $384\times700\times2800$ &  41.12 & 0.78\\
              \\
        0.105& $2.44\times10^5$ & $4.00\times10^2$ & $64\times160\times512$ & 2.907  &  0.10 \\
           0.105& $7.04\times10^5 $& $6.80\times10^2$ & $64\times160\times512$ &  3.653 & 0.26 \\
           0.105& $1.91\times10^6 $ & $1.12\times10^3$ & $64\times192\times768$  & 4.450  & 0.32\\
           0.105& $3.90\times10^6 $ & $1.60\times10^3$ & $128\times256\times1024$ & 5.151  & 0.31\\
           0.105& $9.52\times10^6 $ & $2.50\times10^3$ & $128\times256\times1024$ & 6.096  & 0.28\\
           0.105& $2.39\times10^7 $ & $3.96\times10^3$ & $160\times384\times1536$ &  7.933 & 0.49\\
           0.105& $4.77\times10^7 $ & $5.60\times10^3$ & $160\times400\times1600$ &11.15   & 0.37\\
           0.105& $9.75\times10^7 $ & $8.00\times10^3$ & $192\times512\times2048$ &  15.24 & 0.68\\
           0.105& $2.15\times10^8 $ & $1.19\times10^4$ & $192\times512\times2048$  & 20.32  & 0.42\\
            0.105& $4.62\times10^8 $ & $1.74\times10^4$ & $256\times640\times2560$ &  26.56 & 0.83\\
            0.105& $9.75\times10^8 $ & $2.53\times10^4$ & $256\times640\times2560$ & 34.76  & 0.57\\
             \\
            0.209& $1.03\times10^5$ & $2.60\times10^2$ & $64\times160\times512$ &  2.452 &  0.19 \\
       0.209& $2.44\times10^5$ & $4.00\times10^2$ & $64\times160\times512$ &  3.333 &  0.24 \\
           0.209& $7.04\times10^5 $& $6.80\times10^2$ & $64\times160\times512$ & 4.516  & 0.39\\
           0.209& $1.91\times10^6 $ & $1.12\times10^3$ &$64\times192\times768$ &  5.364 & 0.23\\
           0.209& $3.90\times10^6 $ & $1.60\times10^3$ & $128\times256\times1024$ &  6.489 & 0.47 \\
           0.209& $9.52\times10^6 $ & $2.50\times10^3$ & $128\times256\times1024$ & 8.016  & 0.59\\
           0.209& $2.39\times10^7 $ & $3.96\times10^3$ &$160\times384\times1536$ &  11.73 & 0.63 \\
           0.209& $4.77\times10^7 $ & $5.60\times10^3$ & $160\times400\times1600$ &  14.15 & 0.48\\
           0.209& $9.75\times10^7 $ & $8.00\times10^3$ &$192\times512\times2048$   &  17.93 & 0.72 \\
           0.209& $2.15\times10^8 $ & $1.19\times10^4$ &$192\times512\times2048$   &  22.49 & 0.61\\

  \end{tabular}
  \caption{Values of the control parameters and the numerical results of the simulations. Three series of different groove height are presented. In each series, we vary the $Ta$ and thus the $Re_i$ number. The fourth column shows the amount of grids used in azimuthal ($N_\theta$), radial ($N_r$), and axial direction ($N_z$). The fifth column shows the dimensionless torque, $Nu_\omega$. The last column shows the criterions of resolution we choose, i.e. angular velocity current difference along the radius $J_\omega$. All of the simulations were run in reduced geometry with $L=2\pi/3$ and a rotation symmetry of the order 6. The corresponding cases at the same $Ta$ without grooves (with smooth cylinders) can be found in \cite{ostilla2013,ostilla2014a}.}
  \label{tab1}
  \end{center}
\end{table}

\subsection{Explored phase space}
Inspired by the literature on RB with roughness \citep{shen1996,du2000}, in which it was found that the heat flux increase can not be solely explained by the increase of surface area, three different groove heights with the same total area in each series, i.e. $\delta=0.052d$, $\delta=0.105d$, $\delta=0.209d$, corresponding to 20, 10, 5 grooves on both surfaces of cylinder, were analysed. In each series with the same groove height, $Ta$ is varied from $10^5$ to $10^9$. We then directly compare our results with our prior simulations by \cite{ostilla2013,ostilla2014a} and with experiments by \cite{lewis1999}, all without grooves. The details of all simulations are displayed in table \ref{tab1}.

\section{Global response: dimensionless torque}
\label{sec3}
As mentioned above, the global response of transport of the TC system can be expressed as torque which is required to keep the inner cylinder at a fixed angular velocity. To investigate the effect of the grooved walls, in this section, the dimensionless torque $Nu_\omega$ is presented as a function of $Ta$, i.e. $Nu_\omega=ATa^\beta$, with grooved walls, in comparison with the results for smooth walls. 
\begin{figure}
  \centering

  \begin{minipage}[c]{0.48\textwidth}
    \includegraphics[width=2.5in]{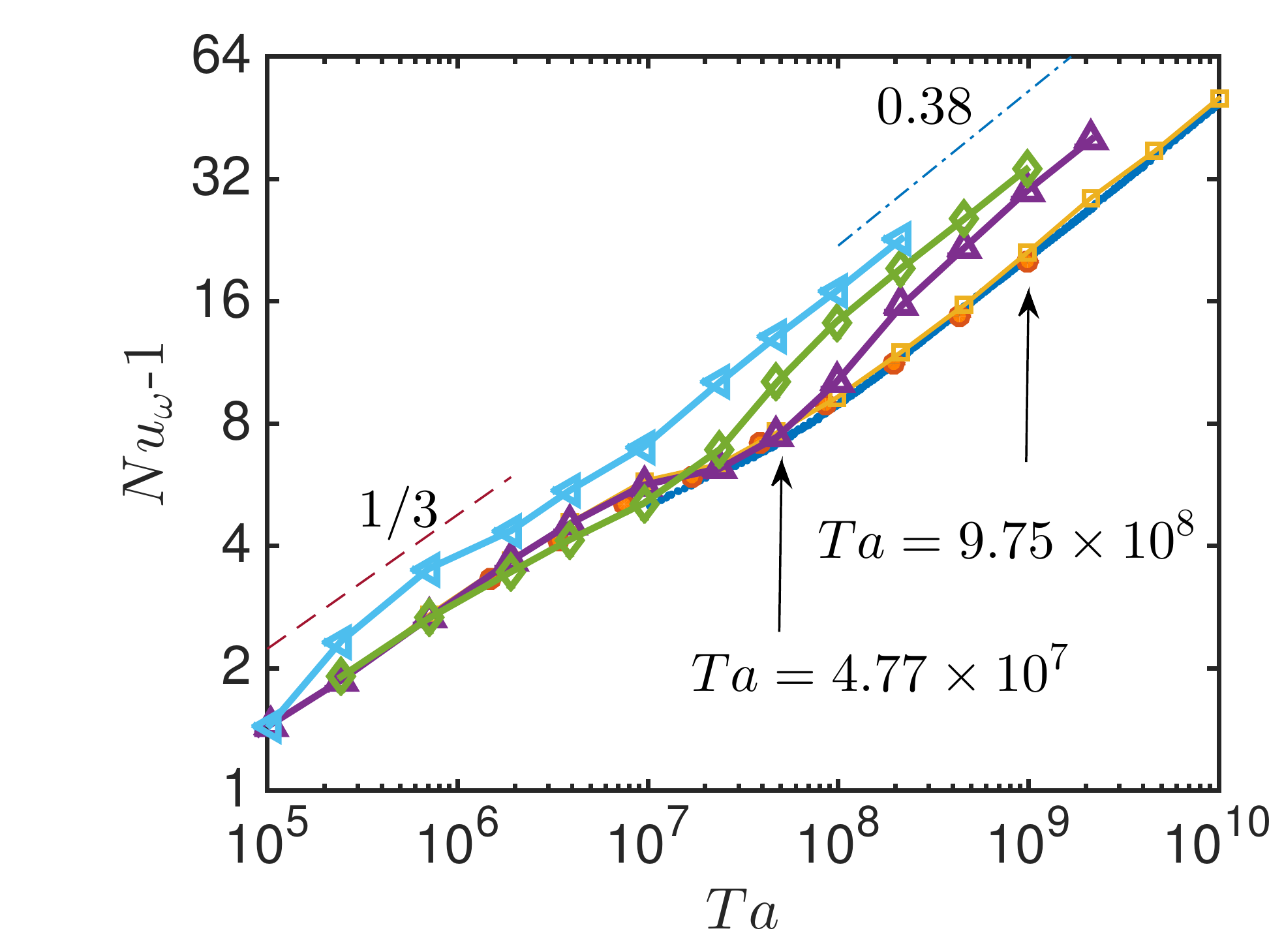}
    \centering {\par (a) \par}
  \end{minipage}%
  \begin{minipage}[c]{0.48\textwidth}
    \includegraphics[width=2.5in]{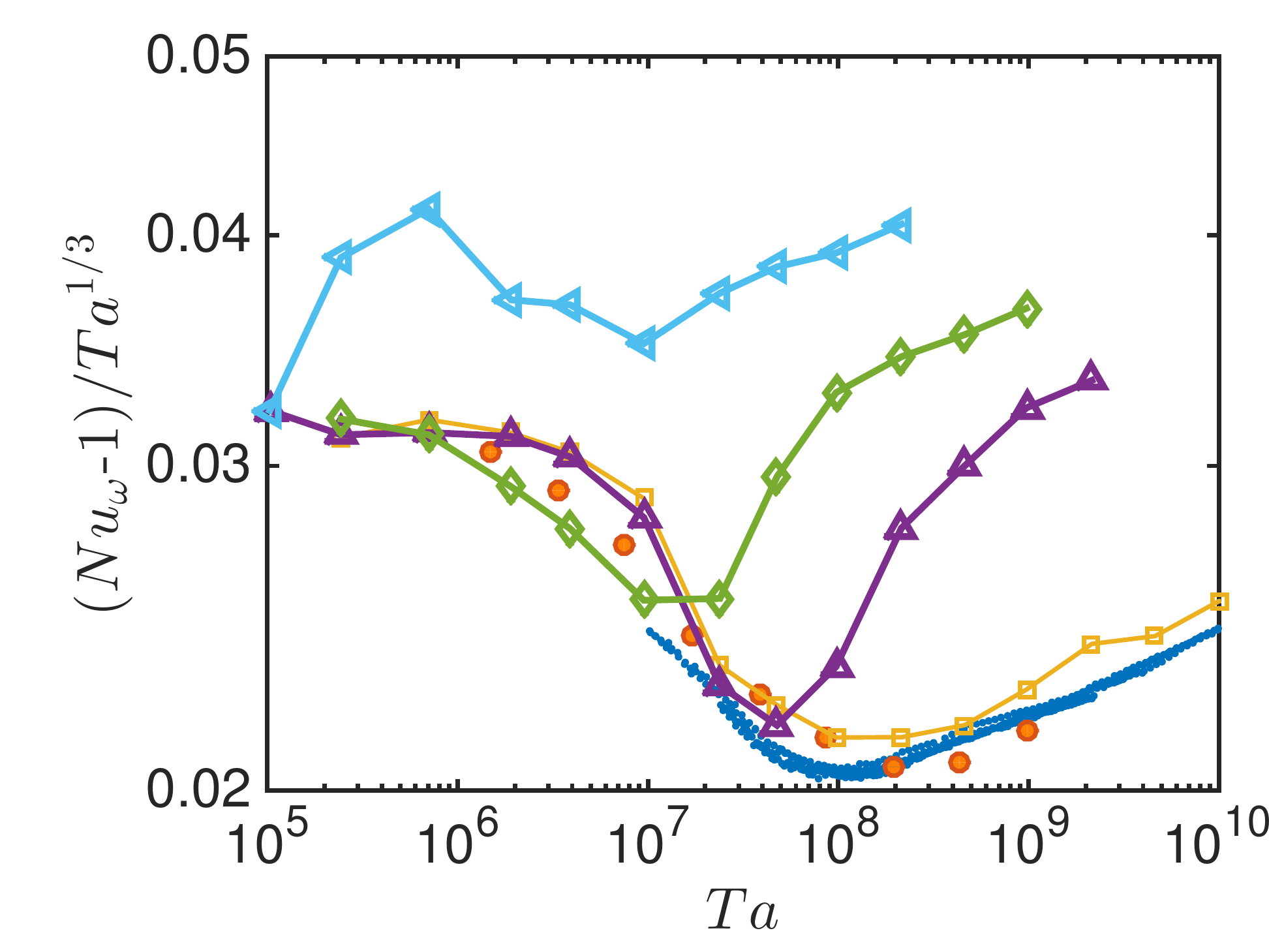}
    \centering {\par (b) \par}
  \end{minipage} 
  \begin{minipage}[c]{0.48\textwidth}
    \includegraphics[width=2.5in]{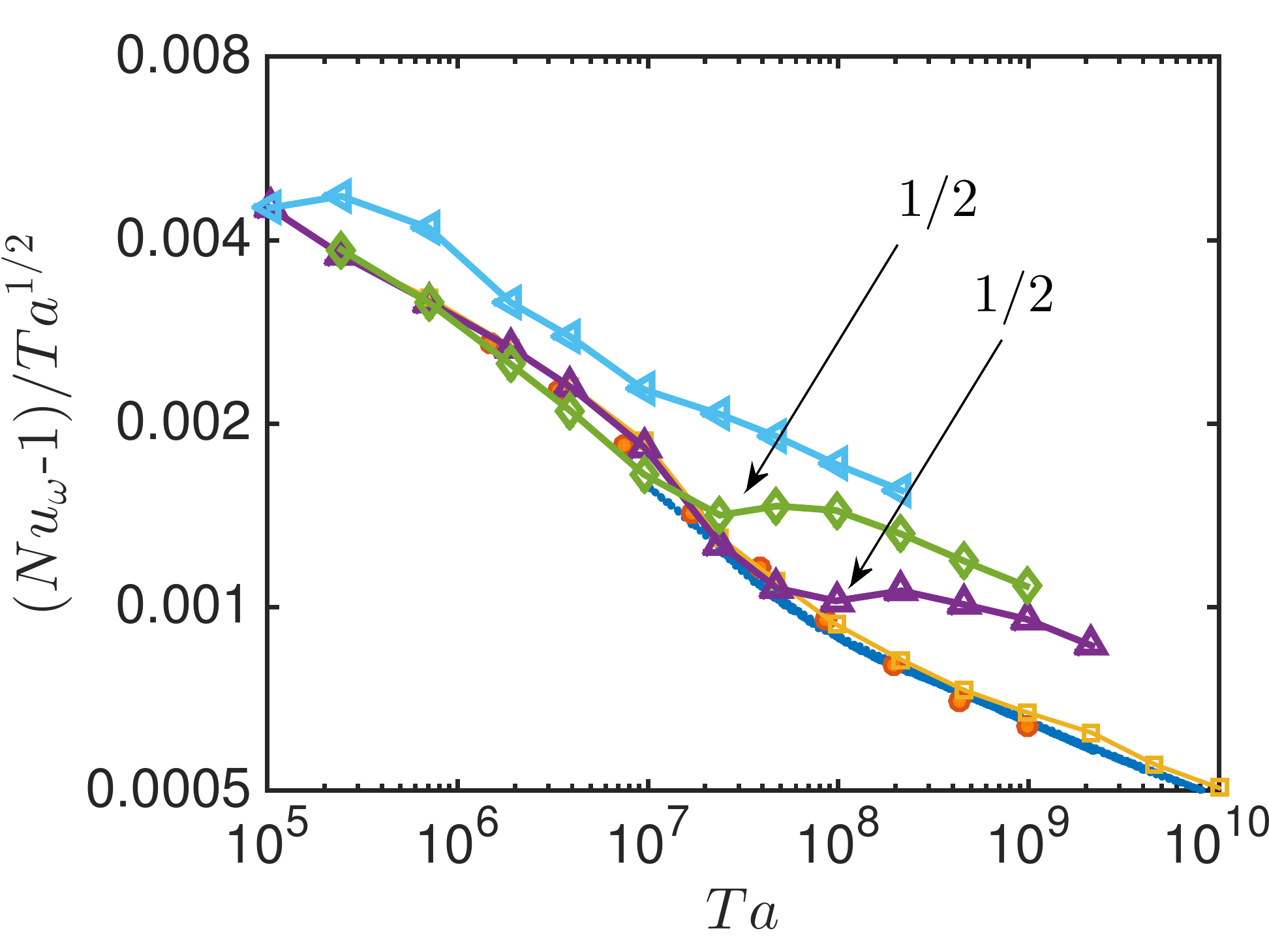}
    \centering {\par (c) \par}
  \end{minipage}
     \begin{minipage}[c]{0.48\textwidth}
    \includegraphics[width=2.5in]{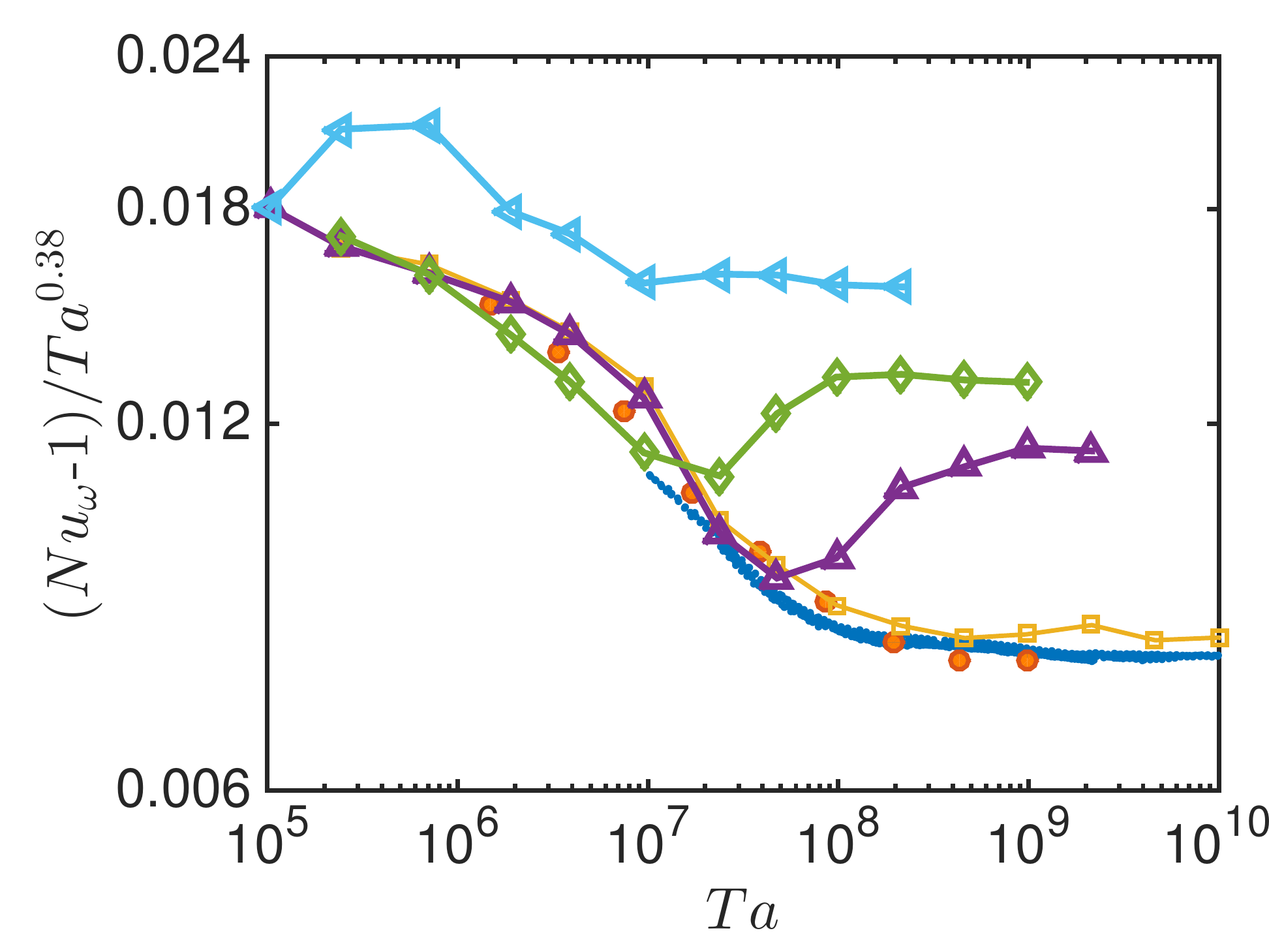}
    \centering {\par (d) \par}
  \end{minipage}
 \caption {(a) Nusselt number as function of $Ta$ for $\eta=0.714$. The data are from experiments and numerical simulations: $\cdot$$\cdot$$\cdot$, smooth walls experiments by \cite{lewis1999}; $\circ$, smooth walls simulations by \cite{braukmann2013}; $\square$, smooth walls simulations by \cite{ostilla2014a}; $\vartriangle$, grooved walls simulations with $\delta=0.052d$ in the present study; $\diamond$, grooved walls simulations with $\delta=0.105d$ in the present study; $\triangleleft$, grooved walls simulations with $\delta=0.209d$ in the present study. Dashed line and dotted dashed  line show the 1/3 and the 0.38 slope. Two arrows direct to the $Ta=4.77\times10^7$ and the $Ta=9.85\times10^8$ at $\delta=0.105d$ series. The former $Ta$ is in the regime where the scaling slope is larger than 0.38 and the latter is in the regime where the slope saturates back to 0.38. We will elaborate these two cases in the following sections as an example to show why the $Nu_\omega$ is increased and why the scaling slope is changed. (b) The same as (a), but now the Nusselt number is compensated with $Ta^{-1/3}$. (c) The same as (a), now with the Nusselt number compensated with $Ta^{-1/2}$. The arrows show the position where local $1/2$ law can be seen. (d) The same as (a), now with the Nusselt number compensated with $Ta^{-0.38}$.}

\label{scaling}
\end{figure}

Figure \ref{scaling} shows $Nu_\omega$ as increasing $Ta$ for smooth cases and grooved cases with three series of different groove heights. For the smooth TC flow, from $Ta=2.5\times10^5$ up to $Ta=3\times10^6$, an effective scaling law of $Nu_\omega \sim Ta^{1/3}$ is found, which is associated with the laminar Taylor vortices. In between $Ta=3\times10^6$ and $Ta=2\times10^8$, there is a transitional region in which first the bulk becomes turbulent, and then the boundary layers also becomes gradually turbulent (see the gradual growing turbulent BL in \cite{ostilla2014a}). When $Ta$ is even larger, the flow is fully turbulent, in both bulk and boundary layer, and the so called ``ultimate'' regime appears with an effective scaling law around $Nu_\omega \sim Ta^{0.38}$ \citep{ostilla2014a,ostilla2014b}. 

The situation is more complicated for the grooved TC flow. As shown in figure \ref{scaling}, for the case of $\delta=0.052d$, three different scaling laws are found with increasing $Ta$: At the early stage, the effective scaling follows the same $Nu_\omega \sim Ta^{1/3}$, as long as $Ta$ is smaller than a threshold Taylor number $Ta_{th}$, at which the effective scaling exponent $\beta$ in the grooved cases starts to deviate from the smooth cases. Once $Ta>Ta_{th}$, $Nu_\omega$  
for the grooved wall cases increases with a steeper exponent $\beta$ than for the smooth wall counterpart. The exponent $\beta$ can locally be 1/2 for this region. However, further increasing $Ta$ leads to a saturation to the ultimate regime effective scaling $Nu_\omega \sim Ta^{0.38}$, already as seen for smooth walls but with a larger prefactor. This means that the influence of the grooves on the effective scaling exponent becomes weaker with increasing $Ta$. 

Note that $Nu_\omega$ does not always increase after $Ta$ exceeds the threshold $Ta_{th}$. For the case of $\delta=0.105d$, at $Ta<Ta_{th}$, the relations between $Nu_\omega$ and $Ta$ of smooth and grooved cases still follow the same route. But after $Ta>Ta_{th}$, there is a region where $Nu_\omega$ is decreased. The largest decrease is 5\% and occurs at $Ta=1.0\times10^7$. The mechanism will be shown and explained in detail in \S \ref{sec4}. This can be related to channel flow in which \cite{choi1993} also found drag reduction in specific range of $Re$. If increasing $Ta$ further, when $Ta>2.5\times10^7$, the exponent $\beta$ starts to increase and can locally be as steep as $1/2$. Again, after that the effective scaling saturates once $Ta$ is large enough. 

For $\delta=0.209d$, the threshold $Ta_{th}$ is at about $Ta=1.0\times10^5$. After a small range of steep regime in which the the slope is larger than 0.38 we can only find the upward shift of $Nu_\omega$. The effective scaling law does not much change, neither in the laminar nor in the turbulent region. This suggests that these massive grooves cannot shift the transition to the ultimate regime to even smaller $Ta$. 

\begin{figure}
\centering
\includegraphics[width=2.5in]{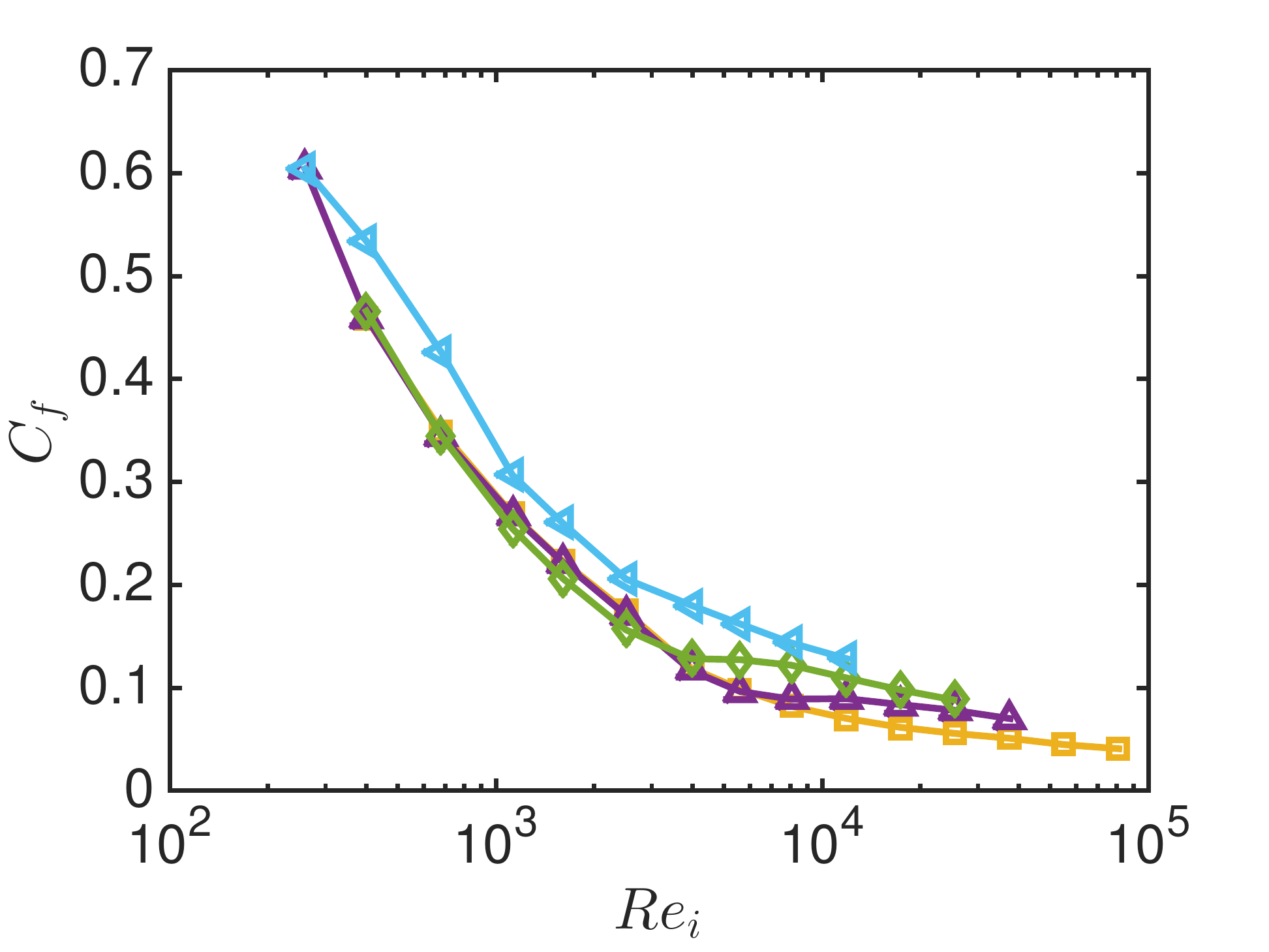}
    \caption{Friction factor $C_f=2\pi Nu_\omega J^\omega_0 \nu^{-2}/Re_i^2$ as a function of the inner cylinder Reynolds number $Re_i$.  $\square$, smooth walls simulations by \cite{ostilla2014a}; $\vartriangle$, grooved walls simulations with $\delta=0.052d$; $\diamond$, grooved walls simulations with $\delta=0.105d$; $\triangleleft$, grooved walls simulations with $\delta=0.209d$.}
\label{cf_re}
\end{figure}

At given $Ta$, we find that a larger groove height causes a more profound increase of $Nu_\omega$. For example, after saturation, for the cases of $\delta=0.052d$, $\delta=0.105d$, and $\delta=0.209d$, $Nu_\omega$ increases by 41\%, 58\%, and 68\%, respectively. The enhanced transport in TC with grooved walls therefore cannot be solely ascribed to the surface area increase, it is rather the local flow dynamics near the grooves which enhances the transport. In the following sections we will look into the flow details to explore the mechanism of $Nu_\omega$ increase which goes beyond the pure increase of surface area by the grooves. Note that the $Nu_\omega$ vs. $Ta$ can also be expressed in terms of friction factor 
\begin{eqnarray}
C_f=2\pi Nu_\omega J^\omega_0 \nu^{-2}/Re_i^2,
\end{eqnarray}
as a function of the inner cylinder Reynolds number $Re_i$, which is shown in figure \ref{cf_re}. From the definition we get that once the local scaling exponent between $Nu_\omega$ and $Ta$ equals to 1/2, a plateau can be found in the $C_f$ vs. $Re_i$ relation.

\begin{figure}
\centering
\includegraphics[width=2.5in]{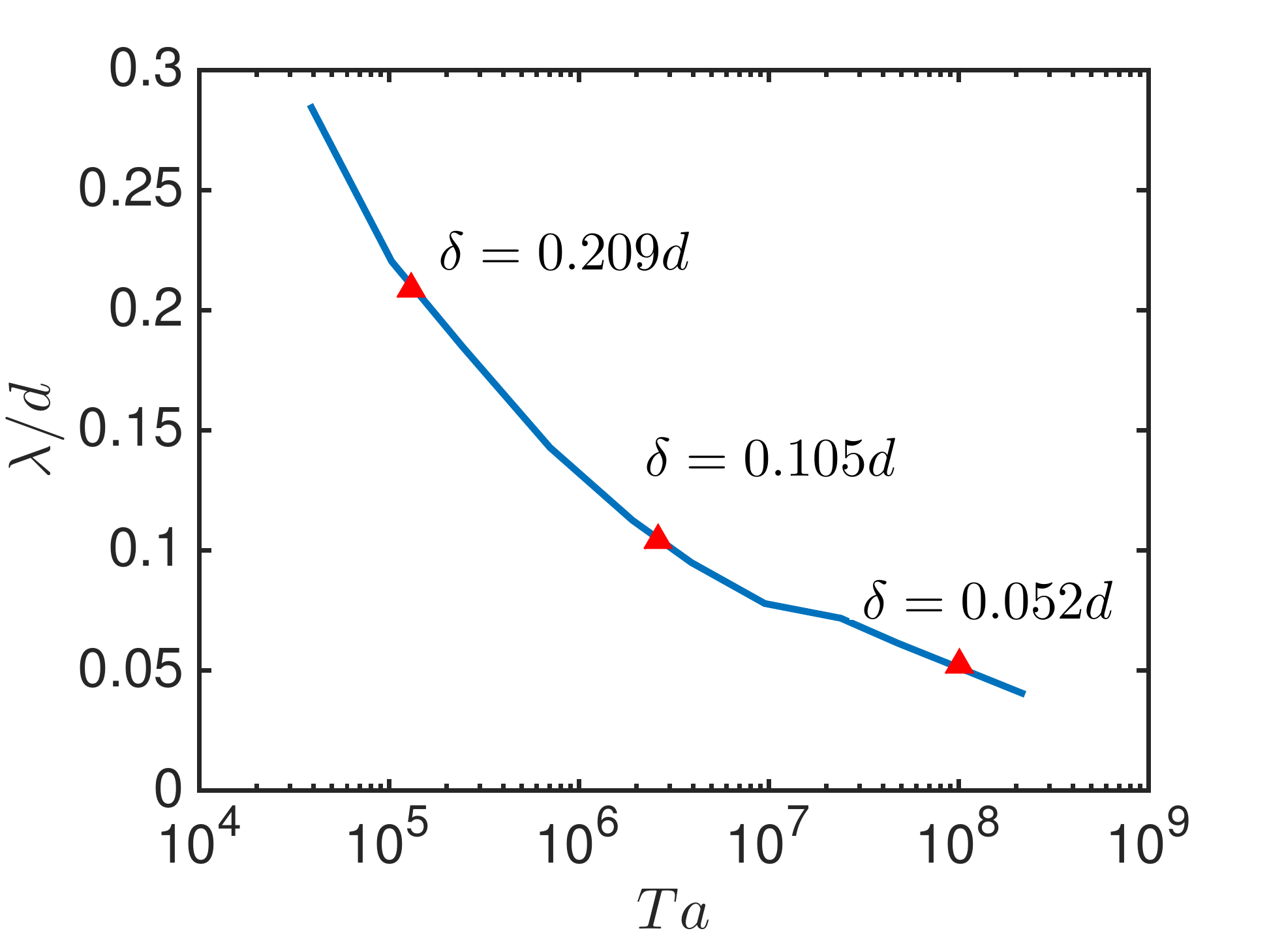}
    \caption{Comparison between boundary layer thickness $\lambda$ and groove height $\delta$. The solid line denotes the averaged boundary layer thickness estimated by $\lambda  \simeq d\sigma/(2Nu_\omega)$ \citep{braukmann2013}. Each triangle symbol shows at what $Ta$ the groove height equals boundary layer thickness.}
\label{BLthickness}
\end{figure}

It is a common finding that in RB flow, the surface roughness becomes active only when the thermal boundary layer is thinner than a representative roughness height \citep{shen1996,stringano2006}. Similar to this, we also want to stress that in TC flow with grooves the effect of grooves on $Nu_{\omega}$ can only be seen when the BL thickness $\lambda$ becomes less than the groove height $\delta$. At small $Ta$, the BL is very thick, thus the grooves are buried under the BL and the fluid cannot feel their influence. At large $Ta$, the BL becomes thinner than the grooves and thus they strongly affect the BL dynamics and thereby alter the transport properties. At this critical value $Ta_{th}$, the BL thickness equals the groove height. The averaged BL thickness $\lambda$ in TC can be estimated by $\lambda  \simeq d\sigma/(2Nu_\omega)$ \citep{braukmann2013}, where $\sigma$ is defined as $\sigma=[(r_i+r_o)/(2\sqrt {r_or_i}))]^4$. Indeed, our simulations show that $Nu_\omega$ starts to change once the boundary layer thickness becomes smaller than the groove height. As shown in figure \ref{BLthickness}, for cases of $\delta=0.209d$, $\delta=0.105d$, and $\delta=0.502d$, we have $Ta_{th}\simeq 1.0 \times 10^5$, $Ta_{th}\simeq 2.8\times 10^6$ and $Ta_{th}\simeq 9.0\times 10^7$, respectively. Also from figure \ref{scaling} it is found that there are sharp transitions for $Nu_\omega$ at these points.

We stress that the local exponent $\beta \approx 1/2$ we have found in figure \ref{scaling} (see the arrows) is not the ultimate region scaling without logarithmic correction but just a crossover between a regime where groove depth is less than the BL thickness and a regime where the groove depth is larger than the BL thickness. Hence it is fortuitous to find this 1/2 exponent in TC with grooves. Note that in pipe flow \citep{nikuradse1933} and TC flow with perpendicular roughness \citep{berg2003} the situation is different. In both cases the roughness is orthogonal to the flow direction and the main flow is impeded by the roughness. The transfer of momentum from the fluid to the wall is accomplished by the drag on the roughness elements, which at high $Re$ is predominantly by pressure
forces, rather than by viscous stresses. A new length scale of the groove height is thus implemented into the system and the drag is independent of $Re$ at high $Re$ \citep{pope2002}. In contrast, in our current simulations with grooves, these are aligned with the flow direction. As a result the grooves do not play an immediate role in generating drag and there is no new length scale to be implemented to the BL. When increasing $Ta$, the viscous stresses still dominate the drag. Therefore the exponent $\beta$ saturates to the same value as for the smooth case at larger $Ta$. It is important to notice that in RB flow the large scale flow fluctuates and
so it always sees the roughness as small obstacles generating direct drag. This aspect is
different from 
 TC flow. However, drag in TC flow is caused by the wall shear rate of the azimuthal velocity.
  This is analogous to  the wall gradient of the temperature in RB flow, not to the drag caused by the 
  velocity field.


From the above findings, we summarize that compared to smooth cases, in grooved wall TC flow there are three different characteristic regimes. First, when $Ta<Ta_{th}$, smooth and groove cases show the same behaviour. Second, when $Ta>Ta_{th}$, the scaling exponent $\beta$ increases to be locally as large as 1/2. Third, when $Ta$ is large enough, there is a saturation regime in which the exponent $\beta$ saturates back to the ultimate region effective scaling law 0.38, but the prefactor of $Nu_\omega$ is increased by a substantial margin.

\begin{figure}
\vspace{0.5cm}
  \centering
    \includegraphics[width=2.5in]{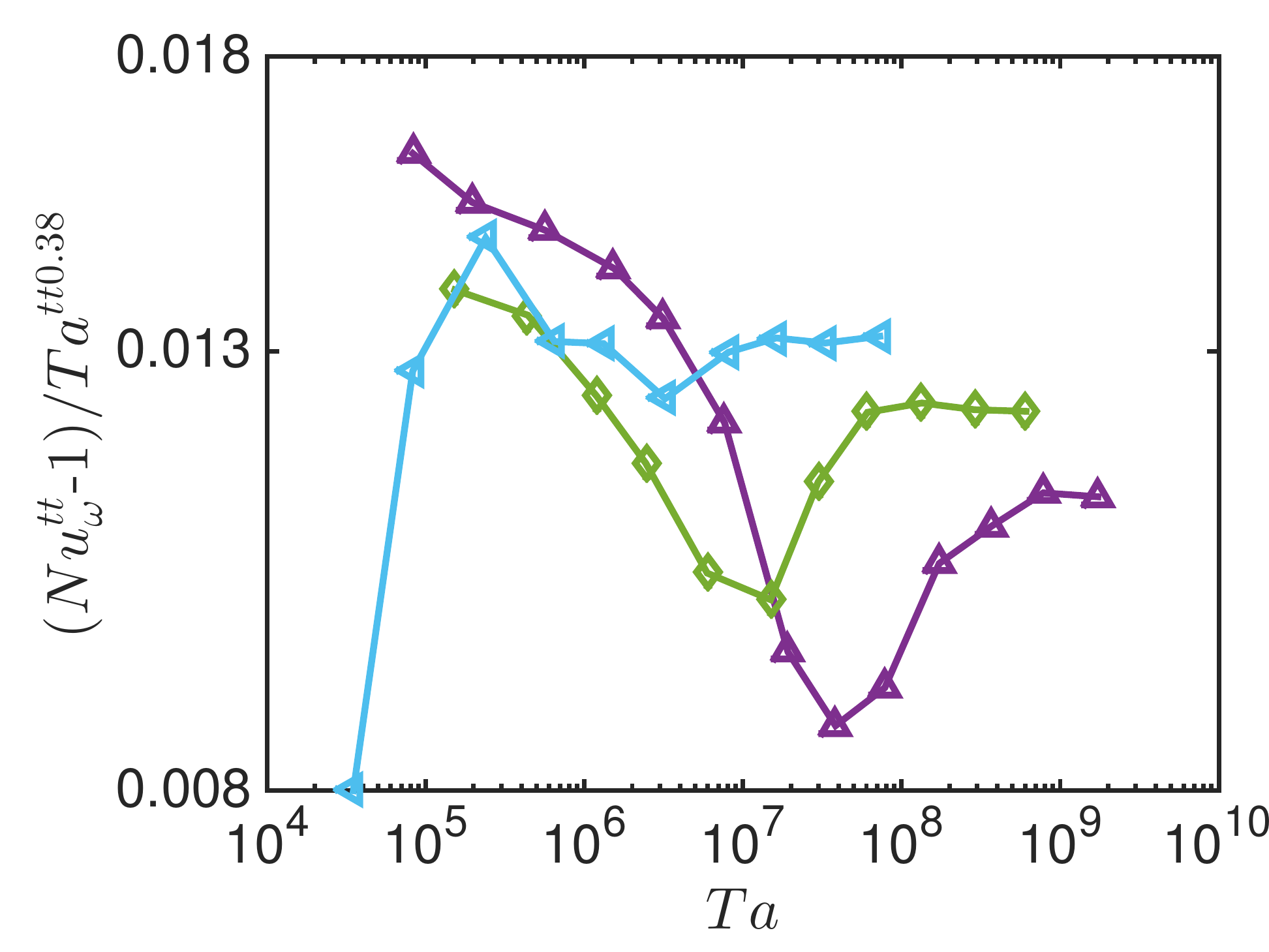}

\caption {Tip to tip Nusselt number as a function of the tip to tip Taylor number compensated with $Ta^{tt-0.38}$ for $\eta=0.714$. $\vartriangle$, grooved walls simulations with $\delta=0.052d$; $\diamond$, grooved walls simulations with $\delta=0.105d$; $\triangleleft$, grooved walls simulations with $\delta=0.209d$. Comparing this figure to the figure \ref{scaling} (d), we find no change in the scaling exponent.}
\label{rescale}
\vspace{0.5cm}
\end{figure}

So far, the $Ta$ and $Nu_\omega$ depicted here are both based on the base to base distance $d$. The usage of $d$ follows the convention of RB studies \citep{du2000,stringano2006}. This choice, however, is quite arbitrary because $d$ is only one of the possibility of the reference length. Also the tip to tip distance could be selected to nondimensionalize $Ta$ and $Nu_\omega$. Such definition would then lead to tip to tip Taylor number $Ta^{tt}=Ta(d-2\delta)^2/d^2$ and Nusselt number $Nu^{tt}_\omega=Nu_\omega(d-2\delta)r_i^2r_o^2/[d(r_i+\delta)^2(r_o-\delta)^2]$. In figure \ref{rescale}, we show the $Nu^{tt}_\omega$ as a function of $Ta^{tt}$. Despite that in our simulations the grooves are quite high, different choices of the characteristic length scale do not affect the effective power laws, but only the exact values of the transitional $Ta$ numbers. Therefore it is reasonable that we only use $d$ as the reference length scale. If the system reached the ultimate state, one would expect that the transport of torque should be the same, i. e. the $Nu_\omega$ vs. $Ta$ relation should be the same as that of the smooth case. When deducting the surface area increase, in our current simulations of $\delta=0.052d$, $\delta=0.105d$, and $\delta=0.209d$, $Nu_\omega$ increases by -1\%, 12\%, and 18\% compared to the smooth case, respectively. We expect that 
   these differences become smaller and smaller with increasing 
    $Ta$. Higher $Ta$ simulations are required to resolve this question.

\section{Flow structures}
\label{sec4}

 \begin{figure}
\centering
\includegraphics[width=5.5in]{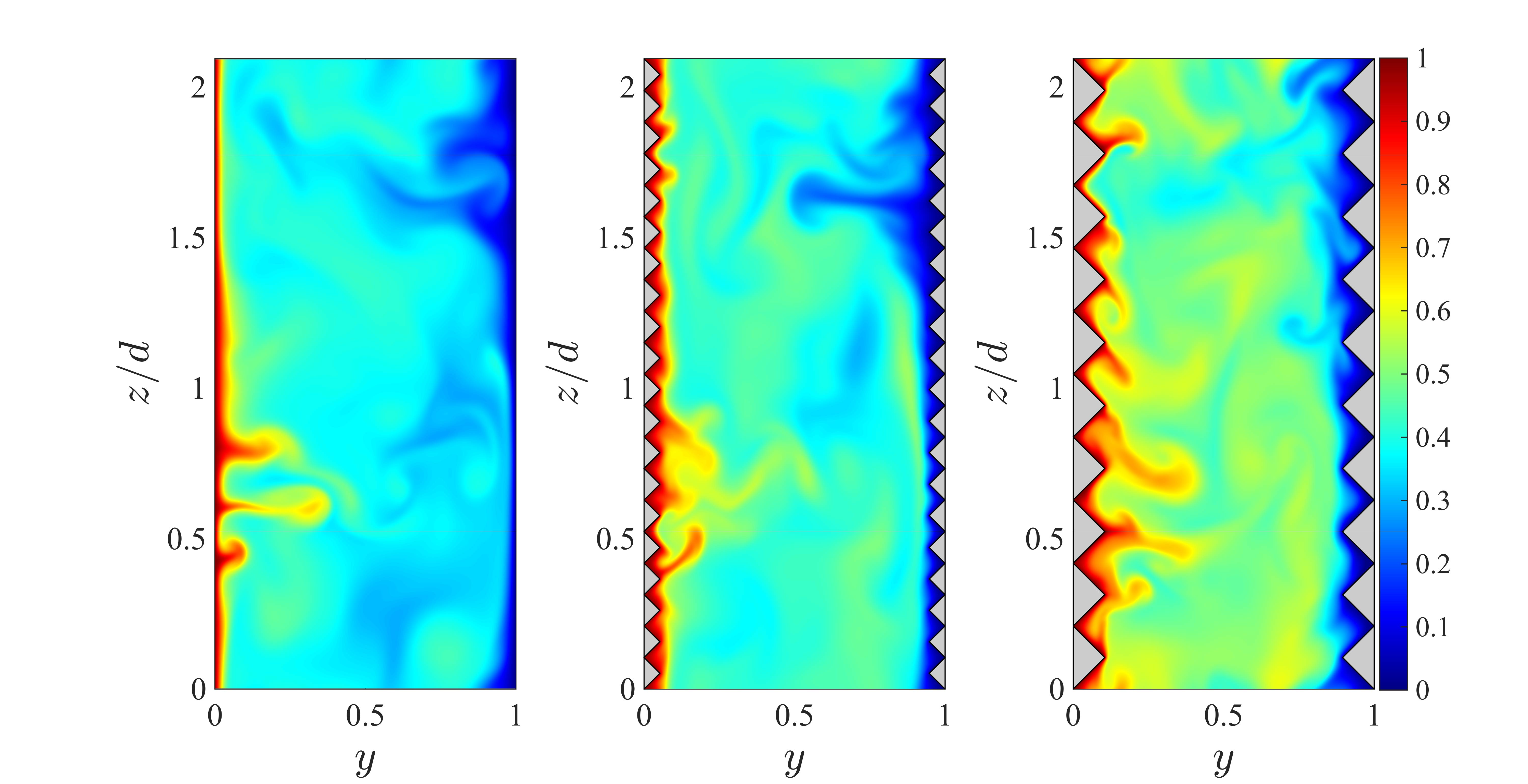}
    \caption{Contour plots of the instantaneous azimuthal velocity field $u_\theta$ for (left panel) the smooth, (middle panel) the $\delta=0.052d$, and (right panel) the $\delta=0.105d$ cases at $Ta=4.77\times10^7$. The colour scale goes from azimuthal velocity 0 to 1. The red colour represents the largest azimuthal velocity 1 while the blue colour represents the smallest azimuthal velocity 0. In the smooth case, the flow is in the transition region where the laminar regions in the boundary layer are coexisting with turbulent, plume ejecting areas. In the middle panel, the boundary layer is still thicker than the groove and therefore no evident difference can be distinguished between the left and middle panel. For the right panel, the boundary layer is thinner than the grooves and plumes are ejected from all tips of the grooves. }
\label{ta_10^7}
\end{figure}

 \begin{figure}
\centering
\includegraphics[width=5.5in]{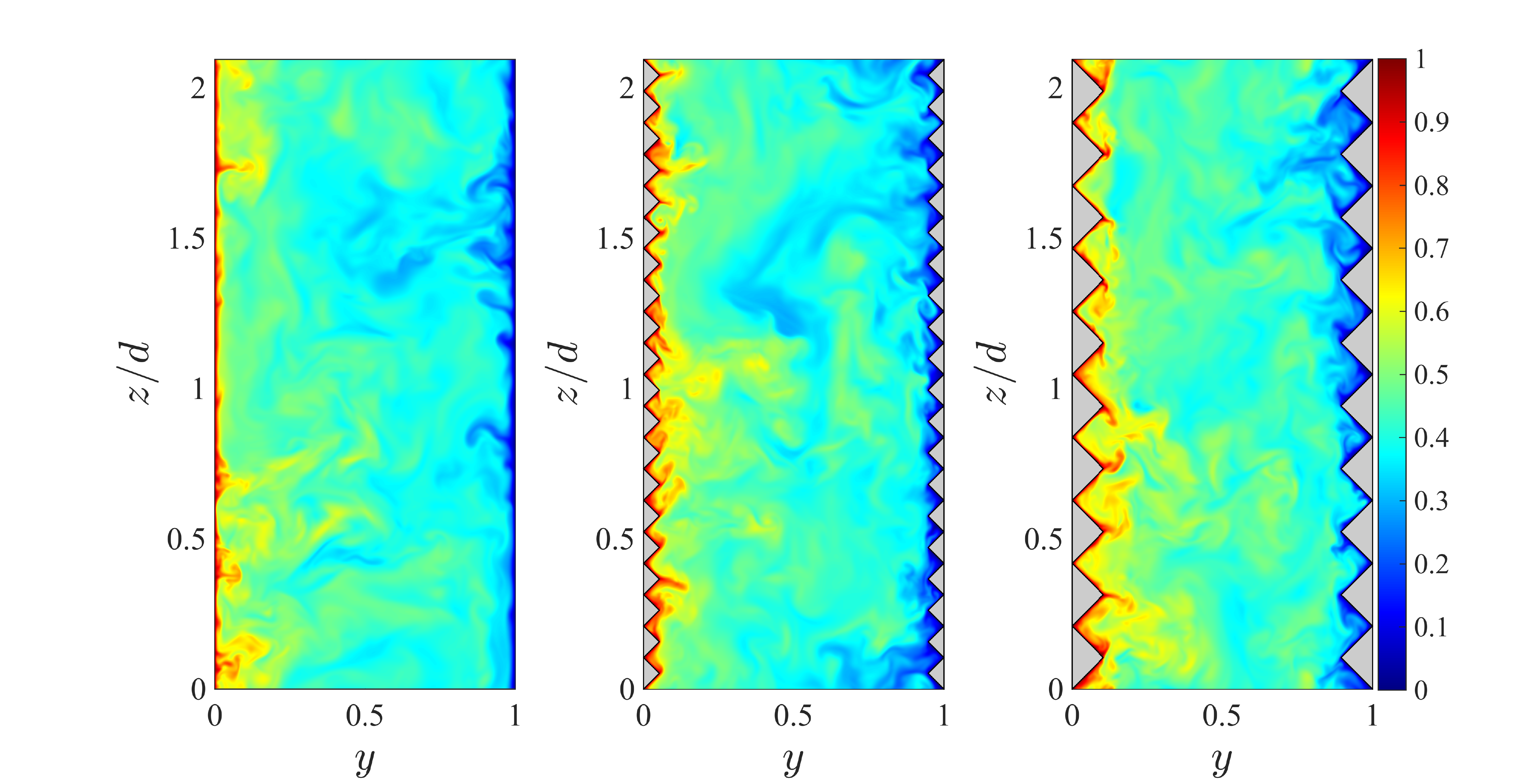}
   \caption{Contour plots of the instantaneous azimuthal velocity field $u_\theta$ for (left panel) the smooth, (middle panel) the $\delta=0.052d$, and (right panel) the $\delta=0.105d$ cases at $Ta=9.75\times10^8$. The colour scale goes from azimuthal velocity 0 to 1. The red colour represents the largest azimuthal velocity 1 while the blue colour represents the smallest azimuthal velocity 0. In the smooth case, the flow is in the fully turbulent region where the bulk and boundary both become turbulent. Due to the decrease of wall pressure gradient in the axial direction, plumes are ejected from more places compared to $Ta=4.77\times10^7$.  In the middle and right panels, where $Ta>Ta_{th}$, plumes are ejected from the tips and these plumes all show preferential directions which follow the Taylor vortices.}
  \label{ta_10^9}
\end{figure}

In order to find the mechanism behind the $Nu_\omega$ increase, in this section, visualizations of the flow in the bulk and the grooves are shown for comparing the grooved and smooth cases. Figure \ref{ta_10^7} shows three contour plots of instantaneous azimuthal velocity $u_\theta$ in a meridional plane for the smooth, the $\delta=0.052d$ and the $\delta=0.105d$ cases at $Ta=4.77\times10^7$. The left panel shows the flow in the transition region where the laminar zones in the BL coexists with turbulent, plume ejecting areas. These plumes are associated with the axial and radial structure which is induced by the Taylor vortices. Plumes are ejected from the preferential positions where there are adverse pressure gradients such that the detachment from the BL is supported. In the middle panel $\delta=0.052d$, $Ta$ is still smaller than $Ta_{th}$, which also means that the grooves are still buried within the BL and hence the effect of the grooves are very small. This is the reason why evident differences cannot be distinguished between the left and the middle panel. However, for the right panel $\delta=0.105d$, $Ta>Ta_{th}$ and the power law between $Nu_\omega$ and $Ta$ which can be seen from figure \ref{scaling} is in the steep regime. We find that plumes are ejected from all the tips of the grooves. It is interesting to note that for the smooth case, the plumes are only ejected from some specific regions while for the grooved case plumes are detached from nearly all tips of the grooves. Because that more plumes are ejected compared to the smooth case, the $Nu_\omega$ is greatly enhanced and the larger exponent $\beta$ regime can be seen in figure \ref{scaling}.

 Figure \ref{ta_10^9} shows three contour plots of instantaneous azimuthal velocity $u_\theta$ in a meridional plane for the smooth, the $\delta=0.052d$ and the $\delta=0.105d$ cases at $Ta=9.75\times10^8$. The left panel shows the flow in the fully turbulent region. Due to the decrease of wall pressure gradient in the axial direction, plumes are ejected from more places compared to $Ta=4.77\times10^7$. For the middle panel $\delta=0.052d$, we have $Ta>Ta_{th}$, and thus we can see that more plumes are ejected due to the existence of grooves. The groove tips are the preferential places for these ejection. In the right panel, for the case of $\delta=0.105d$, compared to the smooth case, plumes are still ejected from the tips of grooves, but near the valleys there are spots where plumes are growing. The difference of the number of plumes between grooved and smooth case is reducing with increasing $Ta$. This maybe another reason why the $Nu_\omega$ versus $Ta$ scaling saturates at high $Ta$. In addition, we note that these plumes have different directions of movement either to the top or bottom. It is interesting to find that these plumes follow the direction of Taylor vortices which also means that Taylor vortices still exist even if there are grooves. This behaviour can be explained by the association between the large scale Taylor vortices and the secondary vortices inside the grooves which we will show in the following paragraphs.
 
%
 
 \begin{figure}
\centering
\includegraphics[width=5.5in]{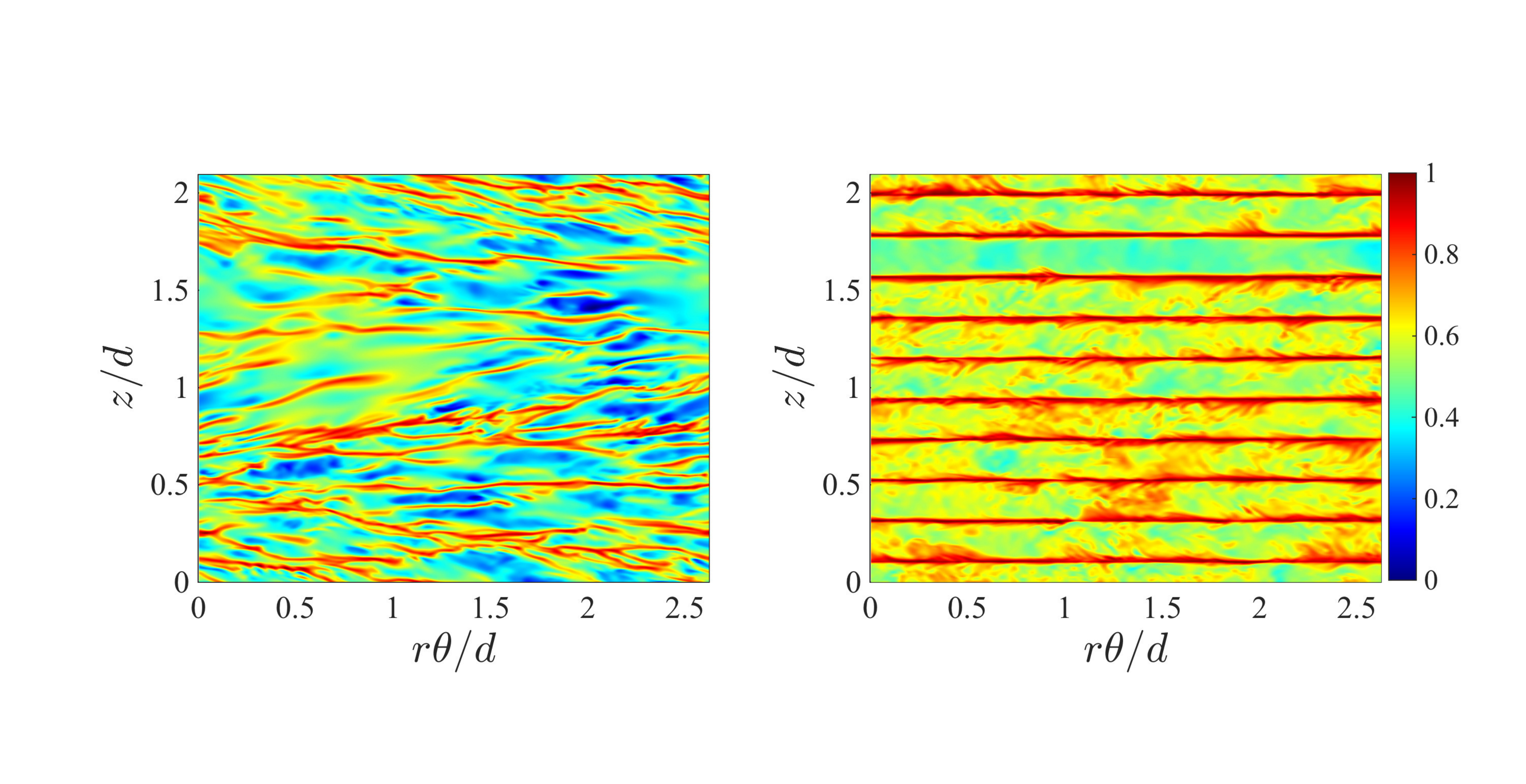}
    \caption{Contour plots of the instantaneous azimuthal velocity $u_\theta(\theta,z)$ for Taylor number $Ta=9.75\times10^8$ at constant radius cuts. The Colour scale goes from azimuthal velocity 0 to 1. The red colour represents the largest azimuthal velocity 1 while the blue colour represents the smallest azimuthal velocity 0. Left panel: the smooth case at fixed wall distance $y=(r-r_i)/d=4\times10^{-3}$. Right panel: the $\delta=0.105d$ case at fixed wall distance in terms of the tip of grooves $y'=(r-r_i-\delta)/d=1\times10^{-3}$. The left panel shows the herring-bone streaks in the boundary layer while the right panel shows plumes being ejected homogeneously along the azimuthal direction.}
\label{radiuscut}
\end{figure}

One may argue that plumes may only be emitted at some points of the groove tips. This however is not the case. Figure \ref{radiuscut} shows contour plots of the instantaneous azimuthal velocity $u_\theta$ of smooth and $\delta=0.105d$ cases for Taylor number $Ta=9.75\times10^8$ at constant radius cuts. In the left panel, at wall distance $y=(r-r_i)/d=4\times10^{-3}$ the flow is in the BL so that the herring-bone streaks can be seen due to the boundary layer instability. While in the right panel, at wall distance $y=(r-r_i)/d=0.106$ or wall distance in terms of the tip of grooves $y'=(r-r_i-\delta)/d=1\times10^{-3}$ the flow is very close to the tips of the grooves, red high speed plumes can be identified almost everywhere near the tip of grooves. It is also seen from this panel, that evidence of large scale vortices can be identified between high speed regions where there are large zones of low speed. This is because that in these regions flows are driven by Taylor rolls and move from outer to inner cylinder or vice versa. In addition, the flow is statistically homogenous in the $\theta$ direction. Both panels can serve as a confirmation for this assumption.

 \begin{figure}
\centering
\includegraphics[width=5.5in]{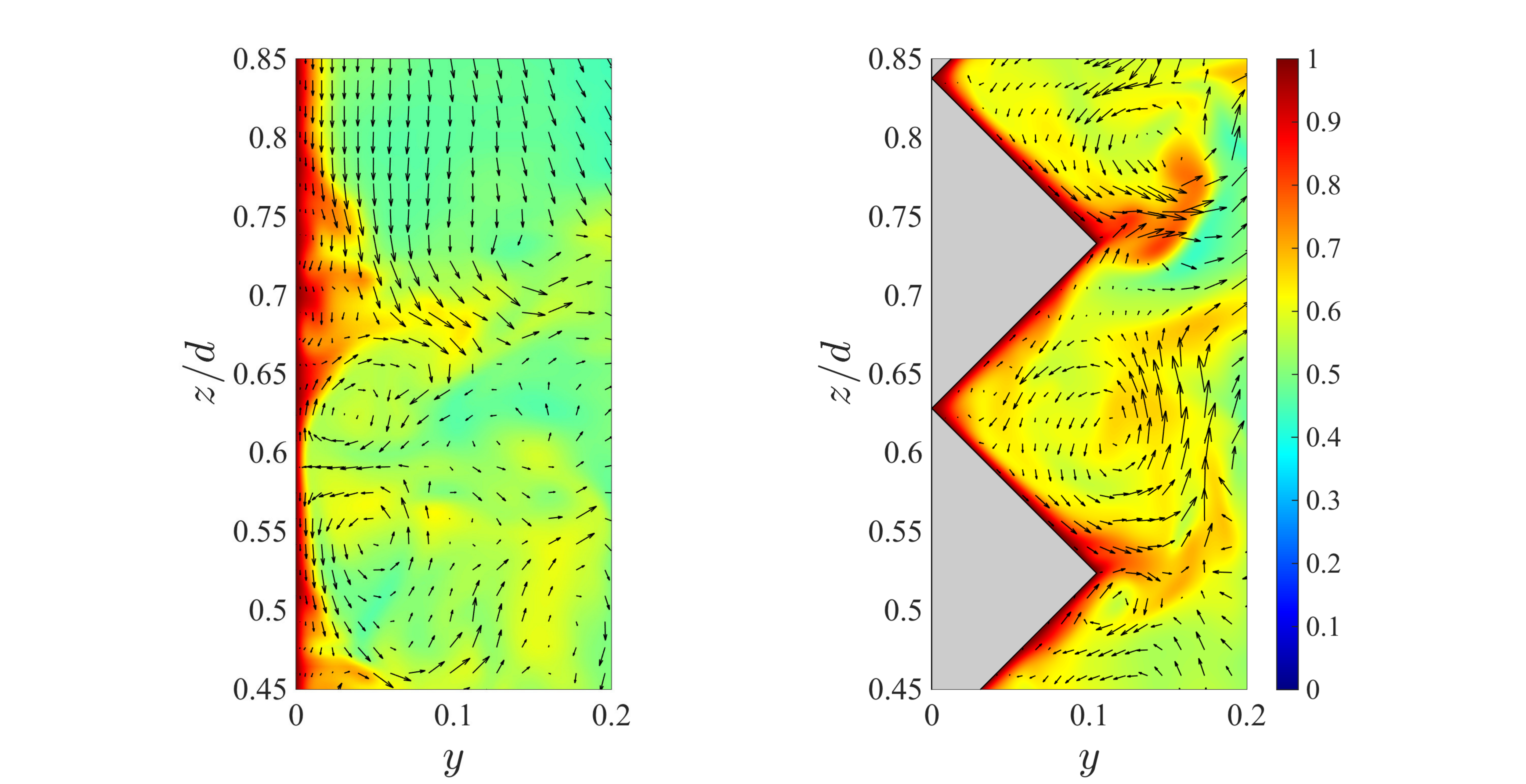}
    \caption{Enlargement of the near wall region for contour plots of the instantaneous azimuthal velocity $u_\theta$ superposed with instantaneous velocity vectors in meridional plane at $Ta=9.75\times10^8$. Left panel: the smooth case. Right panel: the $\delta=0.105d$ case. The colour scale goes from azimuthal velocity 0 to 1. The red colour represents the largest azimuthal velocity 1 while the blue colour represents the smallest azimuthal velocity 0. The velocity vectors are formed by the axial and radial velocity on the surface. For the smooth case in the left panel, the plumes are induced by the acceleration of radial flow. Then vortex rings are formed. For the grooved configuration in the right panel, the axial flow drives the secondary vortex inside the grooves. At the same time the interaction between the secondary vortex and the Taylor rolls causes the detachment of boundary layer from the tips of the grooves. The detached azimuthal flow then develops into a plume.}
\label{streamline}
\end{figure}

We now discuss why plumes are preferentially ejected from the tips of the grooves. Figure \ref{streamline} shows the contour plots of the instantaneous azimuthal velocity $u_\theta$, which is enlarged in plume detaching region of figure \ref{ta_10^9}  and superposed by instantaneous velocity vectors in meridional plane. The left panel displays how plumes detach from the wall of the inner cylinder. Radial pressure gradient accelerates the fluid in the central part of the plume. The sudden acceleration generates secondary vorticity close to the plume and deforms the plume into a mushroom shape. However, for the grooved case, the situation is completely different. The axial pressure gradient favours the Taylor vortex to propel a secondary vortex inside the grooves. This secondary vortex has the opposite direction compared to the large scale Taylor vortex. The interaction between the Taylor vortex and secondary vortex causes the detachment of BL from the tips of the grooves into the bulk and hence forms a plume. Then the plumes are dissipated into the large scale Taylor rolls. These phenomena are clearly seen on the right panel. We note that it is very similar to the famous ``lid-driven-cavity'' flow however it is the Taylor roll which drives the flow in the grooves. Although in RB flows the mechanism for large scale rolls is different from TC, similar thermal plumes are also found to be ejected from the tips of roughness elements there \citep{du2000,stringano2006}.

 \begin{figure}
\centering
\includegraphics[width=5.5in]{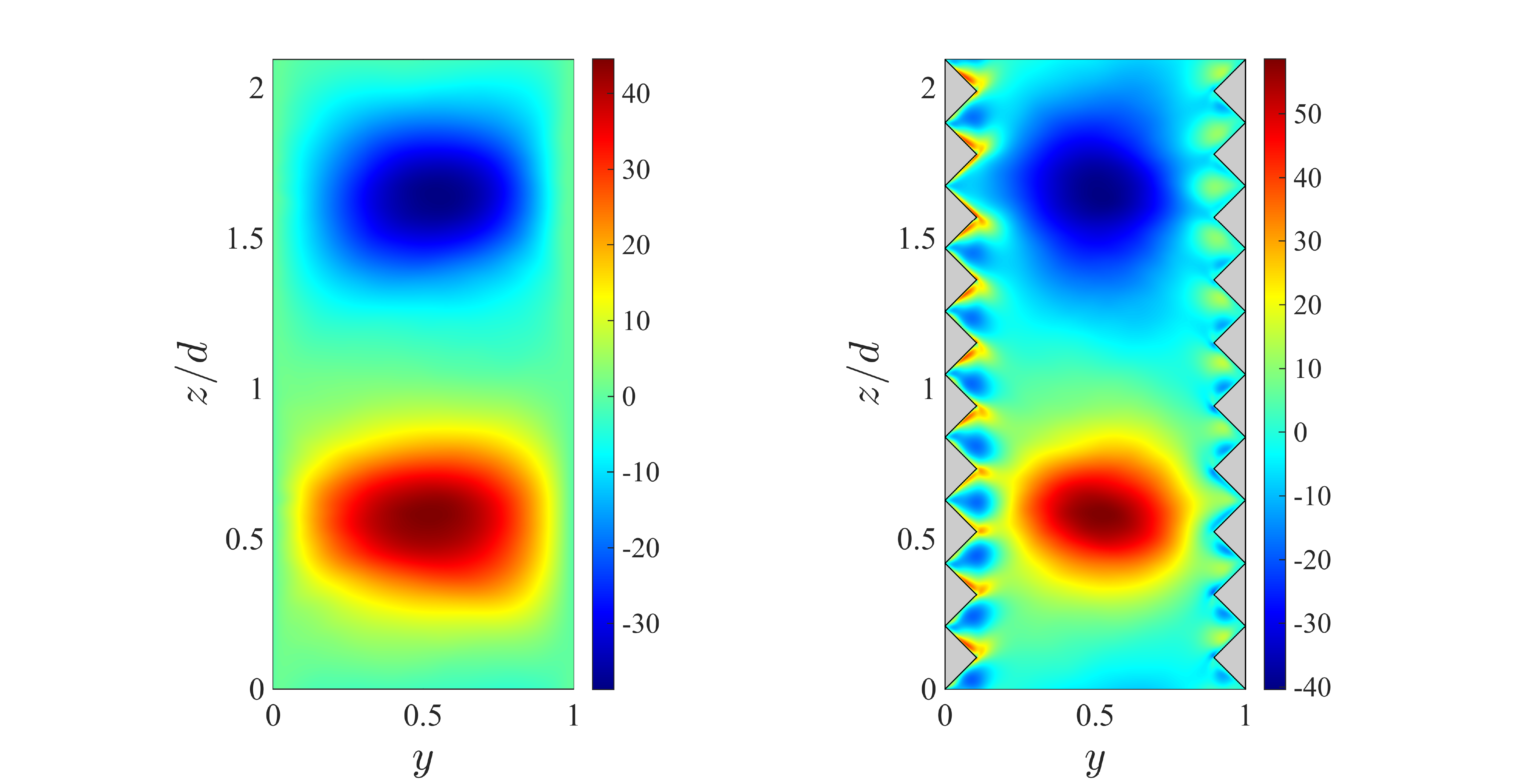}
    \caption{Contour plots of the time- and azimuthally averaged correlation $r^3\langle u_r\omega \rangle_{A,t}$ at $Ta=9.75\times10^8$. Left panel: the smooth case. Right panel: the $\delta=0.105d$ case. The correlation has been normalized according to
 $r^3\langle u_r\omega \rangle_{A,t}/(J_0^{\omega}Nu_\omega)$. The red colour represents the largest value while the blue colour represents the smallest value. It can be seen from the comparison that the interaction between the secondary vortices inside the grooves and large scale Taylor rolls enhances the convective part of the transport in the near tip regions.}
\label{nu_convective}
\end{figure}

It can be seen from figure \ref{streamline} how plumes enhance the $Nu_\omega$ as well. From the definition of angular velocity current $J_\omega$ (equation (\ref{equ1})), it is known that $Nu_\omega$ consists of two parts: convective and conductive contributions. The convective contribution is proportional to the correlation $r^3\langle u_r\omega \rangle_{A,t}$. The interaction between the secondary vortex inside the grooves and Taylor vortex first induce a radial velocity at the tips then lifts the BL. The geometrically induced radial flow separation together with the BL detachment greatly enhances the convective part of Nusselt number. As a evidence for this, in figure \ref{nu_convective}, a comparison of the correlation $r^3\langle u_r\omega \rangle_{A,t}$ at the same $Ta=9.75\times10^8$ between smooth  and $\delta=0.105d$ case is performed. From the right panel it is seen that the BL
is thin enough for the grooves to protrude from it, so that in this case, the interaction of the secondary vortex inside the grooves with the Taylor rolls contributes to the total angular momentum transport while the additional activity in the near tip region determines the extra transport increase. In other words, the grooves enhance plume generation and they allow the plumes to be ejected towards the Taylor rolls. The combined mechanisms then increases the convection part of torque transport.

 \begin{figure}
\centering
\includegraphics[width=5.5in]{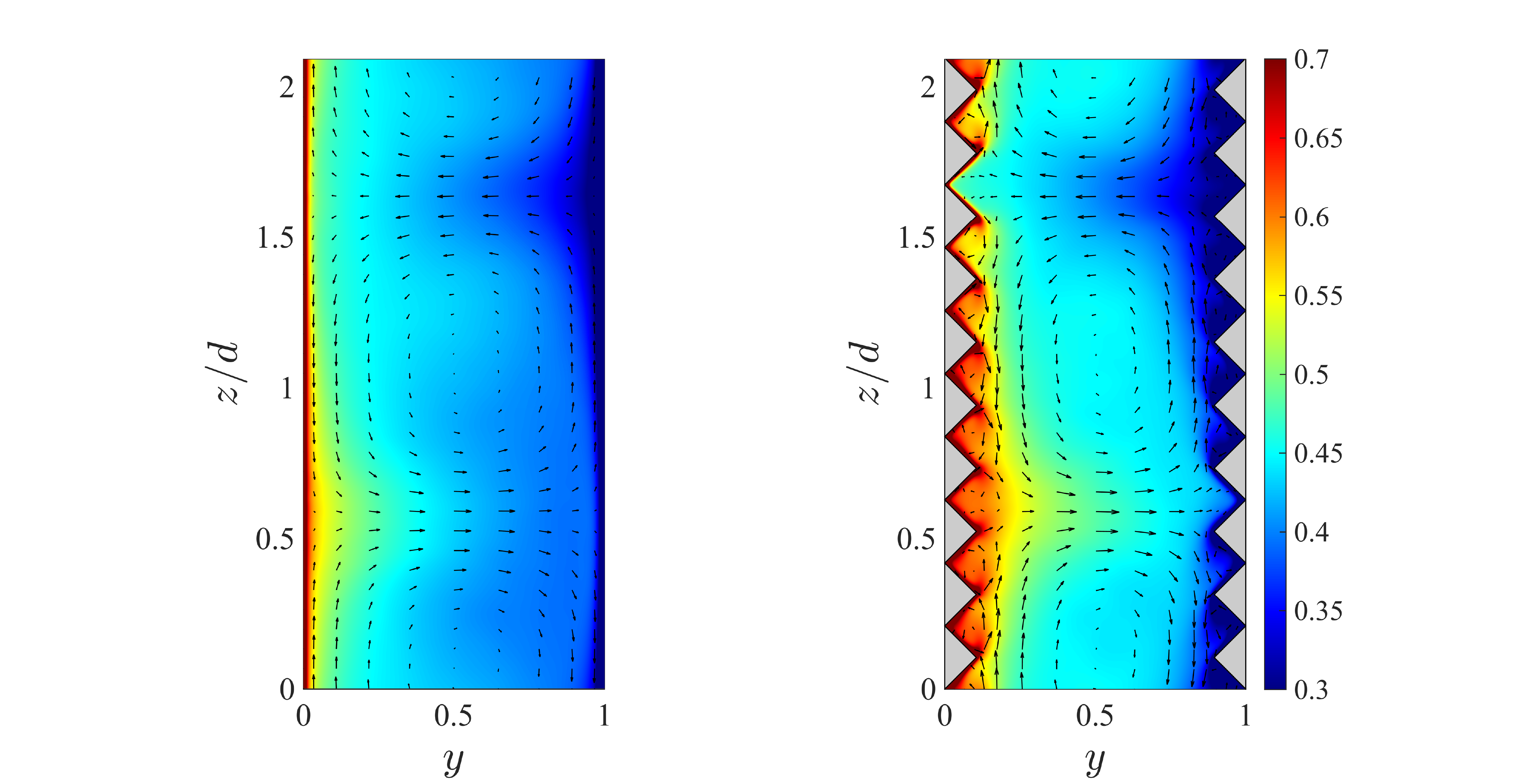}
    \caption{Contour plots of the time- and azimuthally averaged azimuthal velocity $\langle u_\theta \rangle_{\theta,t}$ superposed by time- and azimuthally averaged velocity vectors in meridional plane at $Ta=9.75\times10^8$. Left panel: the smooth case. Right panel: the $\delta=0.105d$ case. The colour scale goes from azimuthal velocity 0.3 to 0.7. The red colour represents the largest azimuthal velocity 0.7 while the blue colour represents the smallest azimuthal velocity 0.3. It can be seen from the comparison that the interaction between the secondary vortices inside the grooves and large scale Taylor rolls favours the circulation of Taylor rolls and hence we get enhanced transport.}
\label{time_averaged}
\end{figure}

From another point of view, we have mentioned before that large scale Taylor vortices still exist in grooved TC flow. In figure \ref{time_averaged}, time-averaged azimuthal velocity contour plots with superposed velocity vectors are presented for the smooth and the $\delta=0.105d$ configurations at $Ta=9.75\times10^8$. The left panel corresponds to a flow field in the ultimate regime. The Taylor roll is still present, but its strength is small and plumes are ejected from many places on the surface. As a comparison, in the right panel it is seen that the large scale Taylor roll interacts with the secondary vortex inside the grooves and therefore we conclude that because of these grooves, on one hand, Taylor roll induces the recirculation inside the the grooves and at the same time the secondary vortex also favours the flow of Taylor rolls more effectively than the smooth case. We define a wind Reynolds number as $Re_w=\sigma(u_r)d/\nu$, where $\sigma(u_r)$ is the standard deviation of the radial velocity. For this figure, we find $Re_w(grooved)=1.12Re_w(smooth)$. In this case the presence of grooves generally leads to a stronger Taylor roll which also contributes to the convective part of torque transport. 
 \begin{figure}
\centering
\includegraphics[width=5.5in]{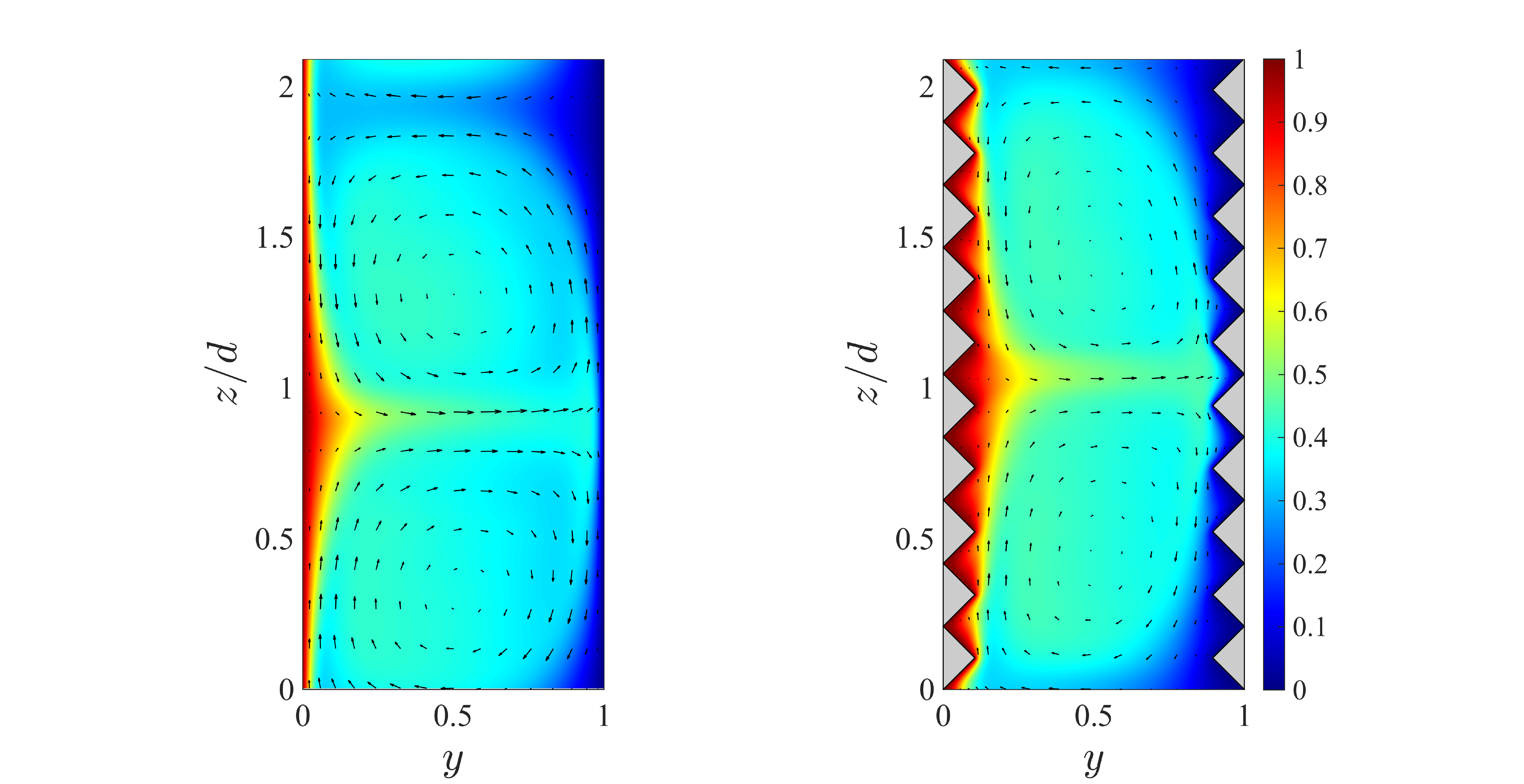}
    \caption{Contour plots of the time- and azimuthally averaged azimuthal velocity $\langle u_\theta \rangle_{\theta,t}$ superposed by time- and azimuthally averaged velocity vectors in meridional plane at $Ta=9.52\times10^6$. Left panel: the smooth case. Right panel: the $\delta=0.105d$ case. The colour scale goes from azimuthal velocity 0 to 1. The red colour represents the largest azimuthal velocity 1 while the blue colour represents the smallest azimuthal velocity 0. Compared to this two panels, no evident secondary vortices can be identified inside the grooves which only serve as obstacles for the Taylor rolls. This is the reason why in some cases we find that $Nu_\omega$ is decreased.}
\label{decreased}
\end{figure}

If, on the contrary, the fluctuation of the Taylor roll is not strong enough to induce the secondary vortex inside the grooves, then stagnant flow inside the grooves could not favour the Taylor vortices. Oppositely, the grooves impede the circulation of Taylor rolls when the rolls flow past the grooves. In figure \ref{decreased}, time-averaged azimuthal velocity contour plot with superposed velocity vectors is presented for the $\delta=0.105d$ configuration at $Ta=9.52\times10^6$. No evident secondary vortices can be seen inside the grooves. Also no large separation of Taylor roll can be identified and these grooves only served as obstacles to hinder the flow to cross them. For this figure, we find $Re_w(grooved)=0.84Re_w(smooth)$. As a result, weaker Taylor vortices deduct the radial angular velocity transport and $Nu_\omega$ is decreased. 

 \begin{figure}
\centering
\includegraphics[width=2.in]{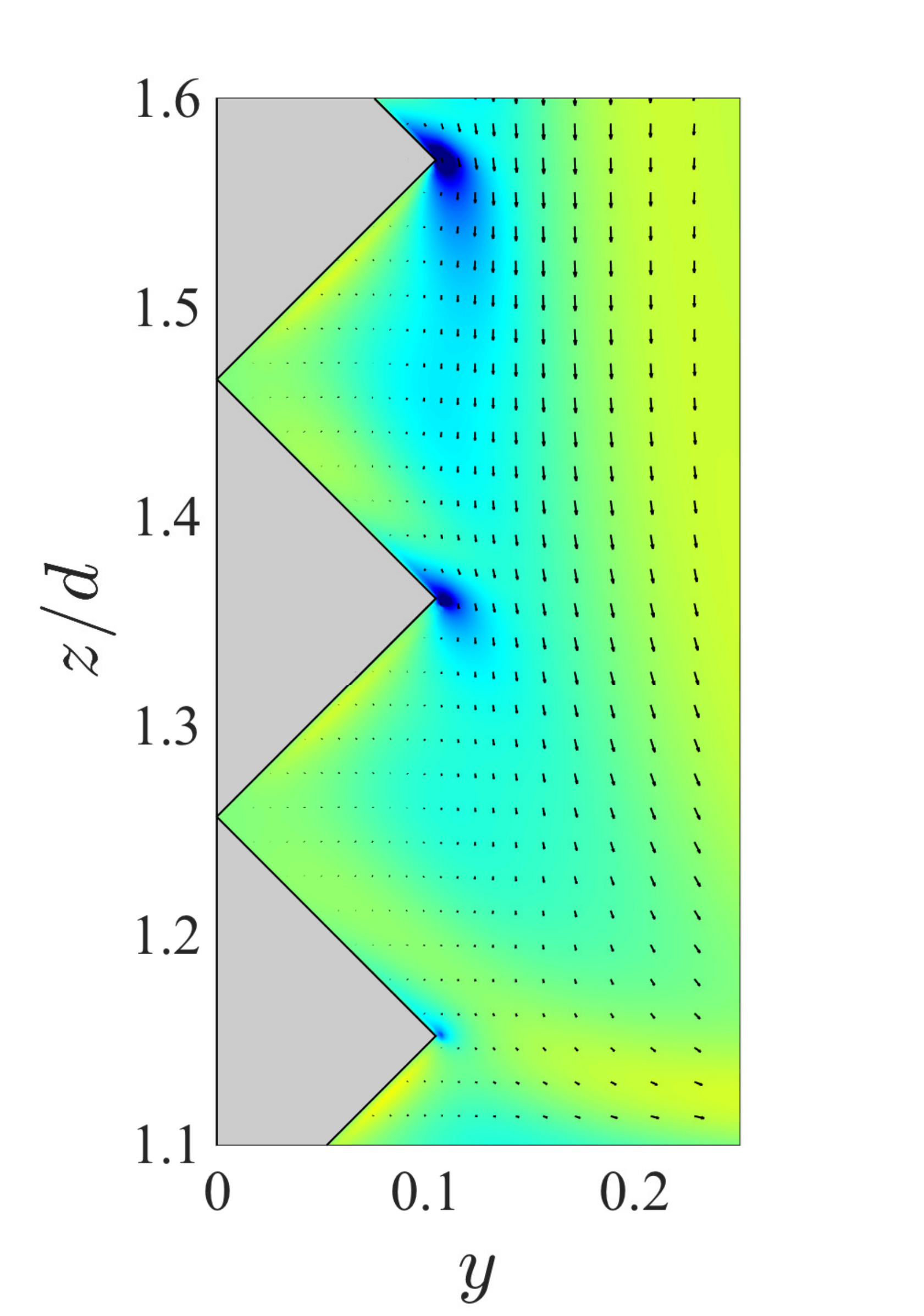}
\includegraphics[width=2.in]{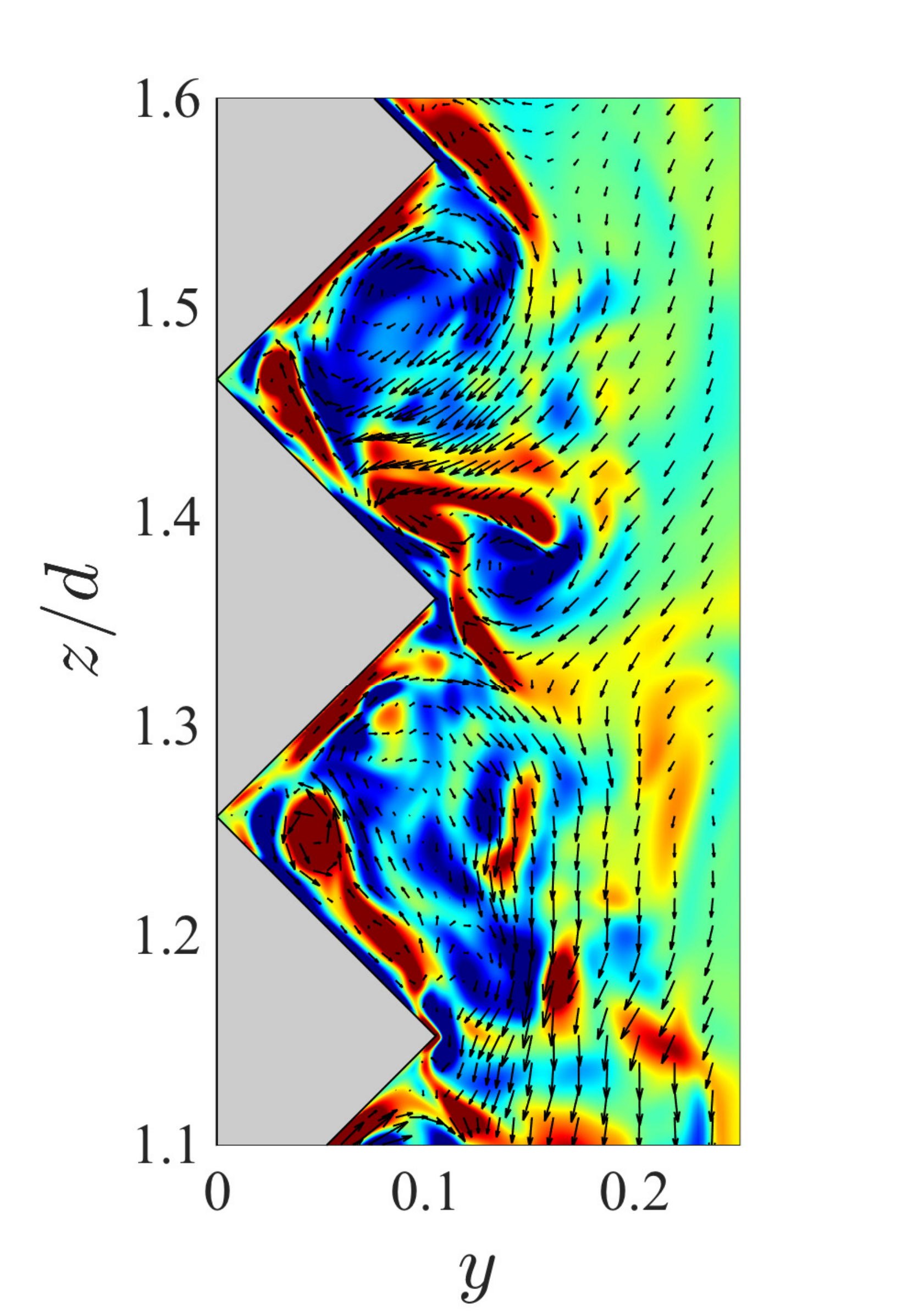}
    \caption{Enlargement of the near wall region for contour plots of the instantaneous streamwise vorticity field superposed with instantaneous velocity vectors in meridional plane at
$Ta = 9.52\times10^6$ (left panel) and $Ta = 9.75\times10^8$ (right panel) with $\delta=0.105d$. The black arrows show the velocity vectors in the plane. The Blue color denotes the negative streamwise vorticity and red the positive.}
\label{vor}
\end{figure}

In the above paragraphs, we discussed how grooves affect the bulk flow and the convective part of $Nu_\omega$. Because of the conservation of angular velocity current $J^\omega$ along the radius, the grooves will also have impact on the wall turbulence structure and wall shear rate. To further illustrate the $Nu_\omega$ increase or decrease mechanism, here we include the instantaneous streamwise (azimuthal) vorticity field snapshot, as shown in figure \ref{vor} at $Ta=9.52\times10^6$ and $Ta=9.75\times10^8$ with $\delta=0.105d$. This figure illustrates the streamwise vorticity field superposed by the velocity vectors in the plane. At $Ta=9.52\times10^6$, there is almost no secondary vortex between the grooves, and only at the tips of the grooves we see the streamwise vortices which are caused by flow separation there. At $Ta=9.75\times10^8$, the secondary vortex is much stronger compared to that of the $Ta=9.52\times10^6$ case. Between groove tips, almost all surface areas of grooves are exposed to the sweep motion that streamwise vorticity (secondary vortex) induces. The stronger secondary vortex is, the more surface area is exposed to the sweep motion it induces, and hence the higher shear rate of azimuthal velocity. We refer to figure \ref{BL} and figure \ref{sv} on how secondary vortices gradually occupy the whole regions between grooves and hence increase the shear rate not only at the tips but also in the valleys with increasing $Re$.

 \begin{figure}
\centering
\begin{minipage}[c]{0.48\textwidth}
      \includegraphics[width=2.5in]{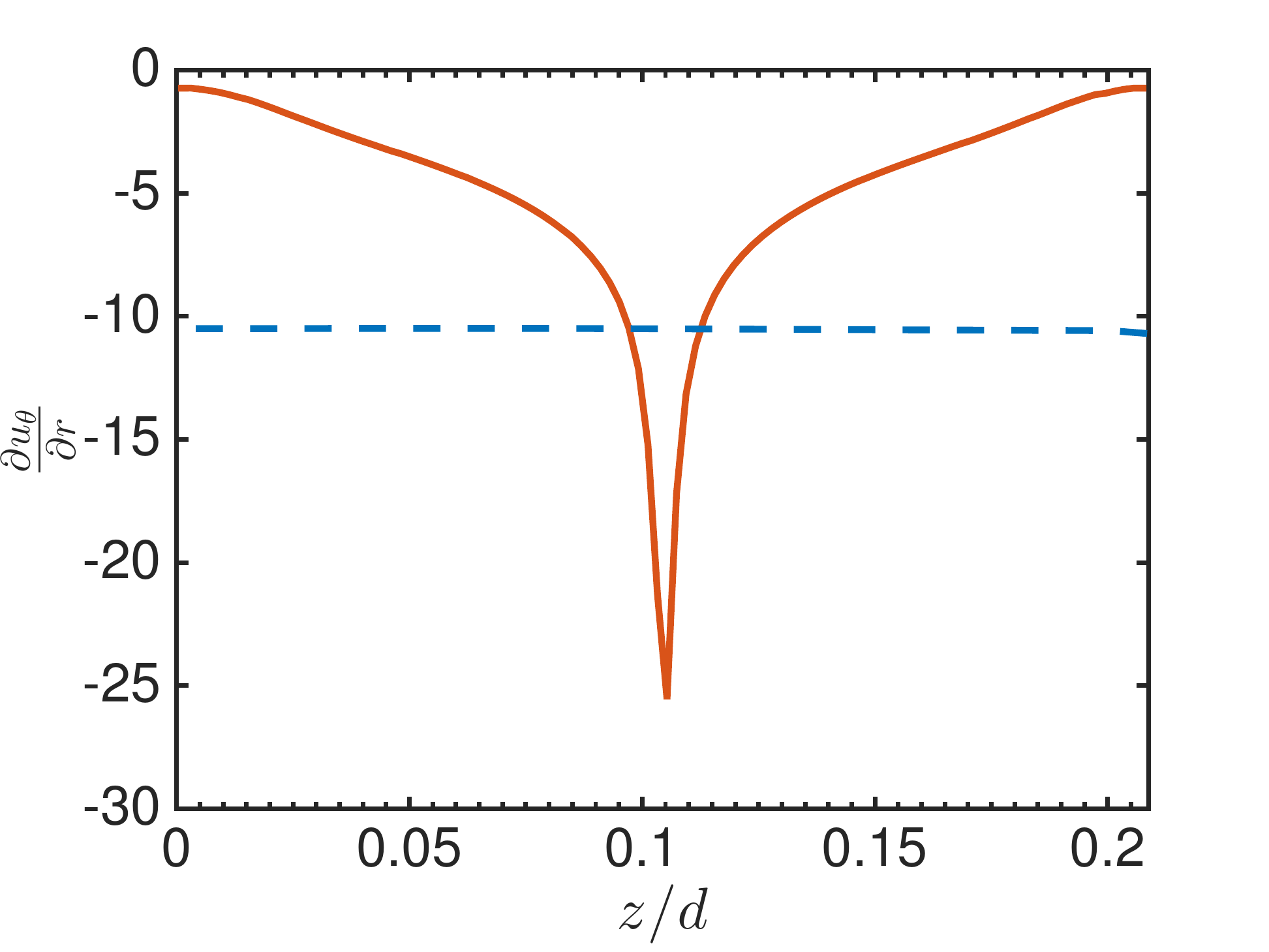}
    \centering {\par (a) \par}
  \end{minipage}%
     \begin{minipage}[c]{0.48\textwidth}
    \includegraphics[width=2.5in]{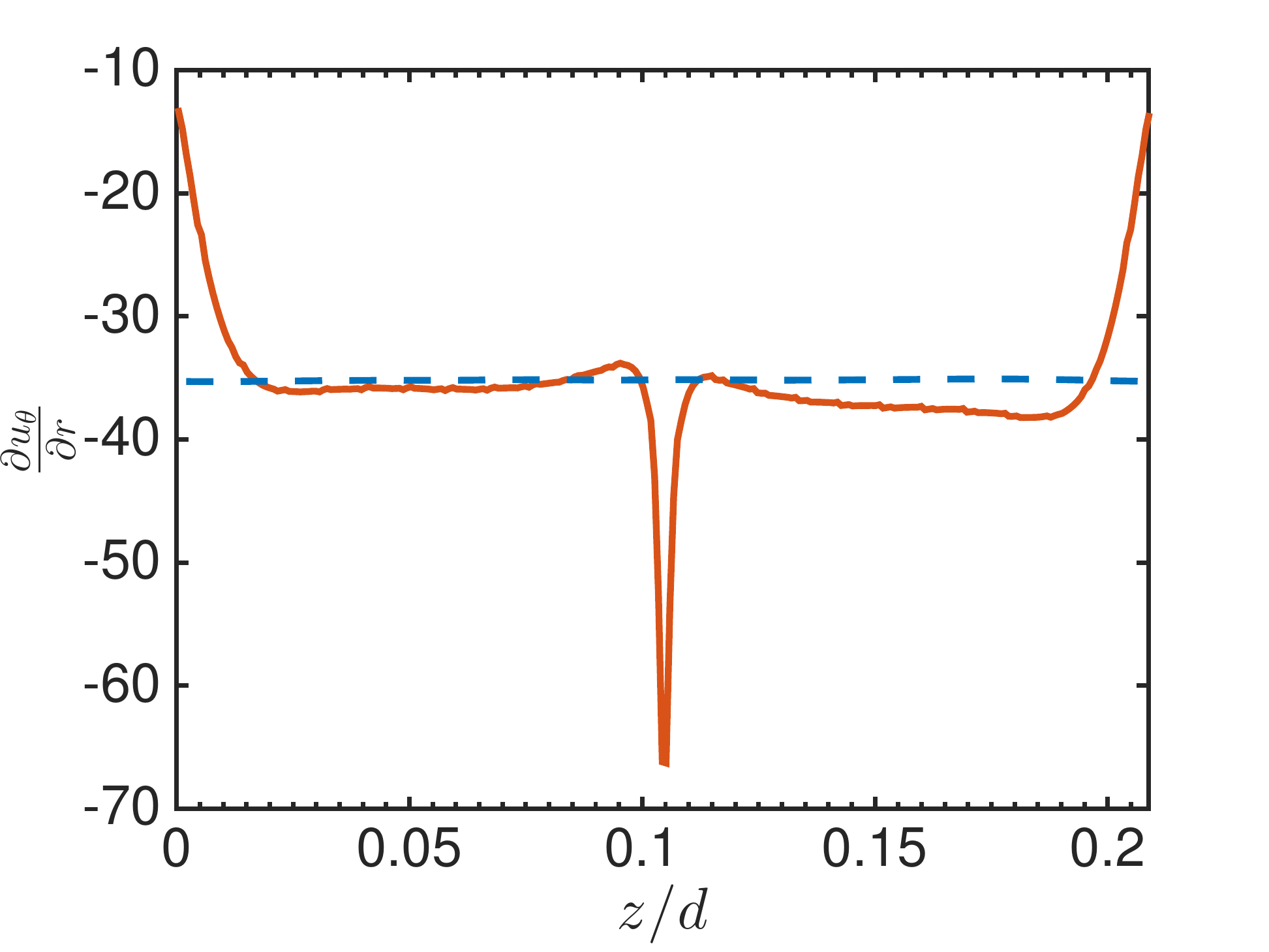}
    \centering {\par (b) \par}
  \end{minipage}
    \caption{Wall shear rate $\frac{\partial u_\theta}{\partial r}$ averaged in time, azimuthally, and over ten grooves for the $\delta=0.105d$ case at two different Taylor numbers: (a) $Ta=9.52 \times10^6$; (b) $Ta=9.75\times10^8$. The solid lines denote the grooved case and the dashed line the averaged wall shear rate for the smooth case at the same $Ta$.}
\label{shear}
\end{figure}  

We now turn to how the secondary vortices are related to the wall shear rate of the azimuthal velocity and hence $Nu_\omega$. In figure \ref{shear}, we show the azimuthal shear rate averaged in time, azimuthally, and over ten grooves at two different Taylor numbers $Ta=9.52\times10^6$ and $Ta=9.75\times10^8$ compared the results with the smooth case. At $Ta=9.52\times10^6$, without secondary vortices, only the tips of the grooves are exposed to the shear of the Taylor rolls and the shear rate decays very quickly to zero from the tips to the valleys of the grooves. There is just a small region close to the groove tip, where the shear rate is larger than that in the smooth case. That is why the torque is reduced. However, at $Ta=9.75\times10^8$, with strong secondary vortices, the strong mixing effect renders almost the whole surface areas to be exposed to the high shear rate equally. Only very close to the valley, one can find that the shear rate is smaller than its smooth counterpart. That is why the torque is enhanced. The mechanism proposed here is similar to the one in channel flow with riblets, where the location of quasi-streamwise vortices is changed by the size of groove and thus turbulence drag increases or decreases according to the groove size \citep{choi1993}. However, we only observe very few cases at one groove height $0.1d$ with $Ta$ number around $10^7$ where torque is reduced. At this $Ta$ the flow is not fully turbulent and we would restrict our analysis to the specific cases. Indeed more work is needed to explore whether drag reduction is possible for fully turbulent TC flow.
 
\section{Boundary layer dynamics}
\label{sec5}

\begin{figure}
  \centering
  \begin{minipage}[c]{0.48\textwidth}
    \includegraphics[width=2.5in]{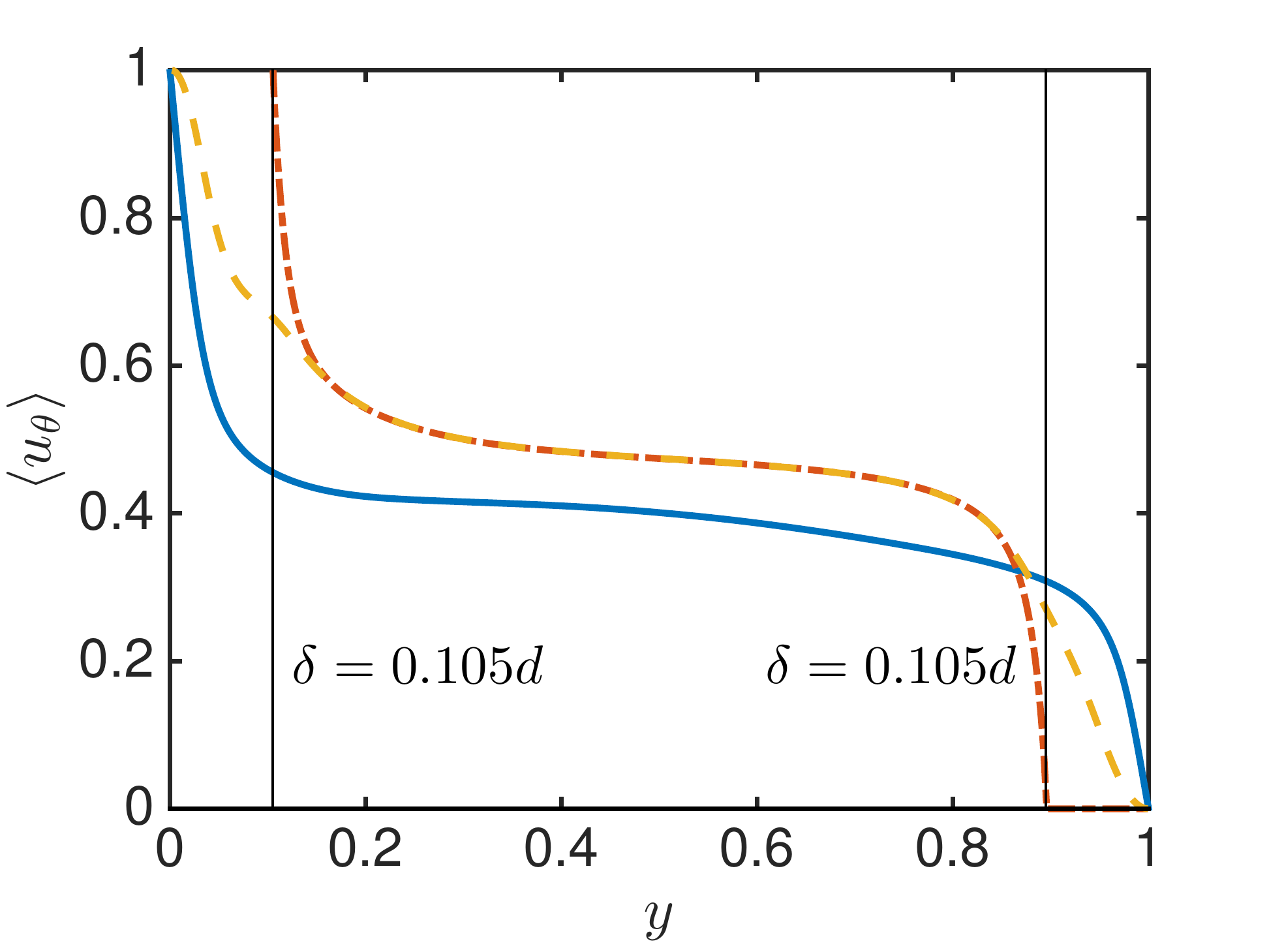}
    \centering {\par (a) \par}
  \end{minipage}%
     \begin{minipage}[c]{0.48\textwidth}
    \includegraphics[width=2.5in]{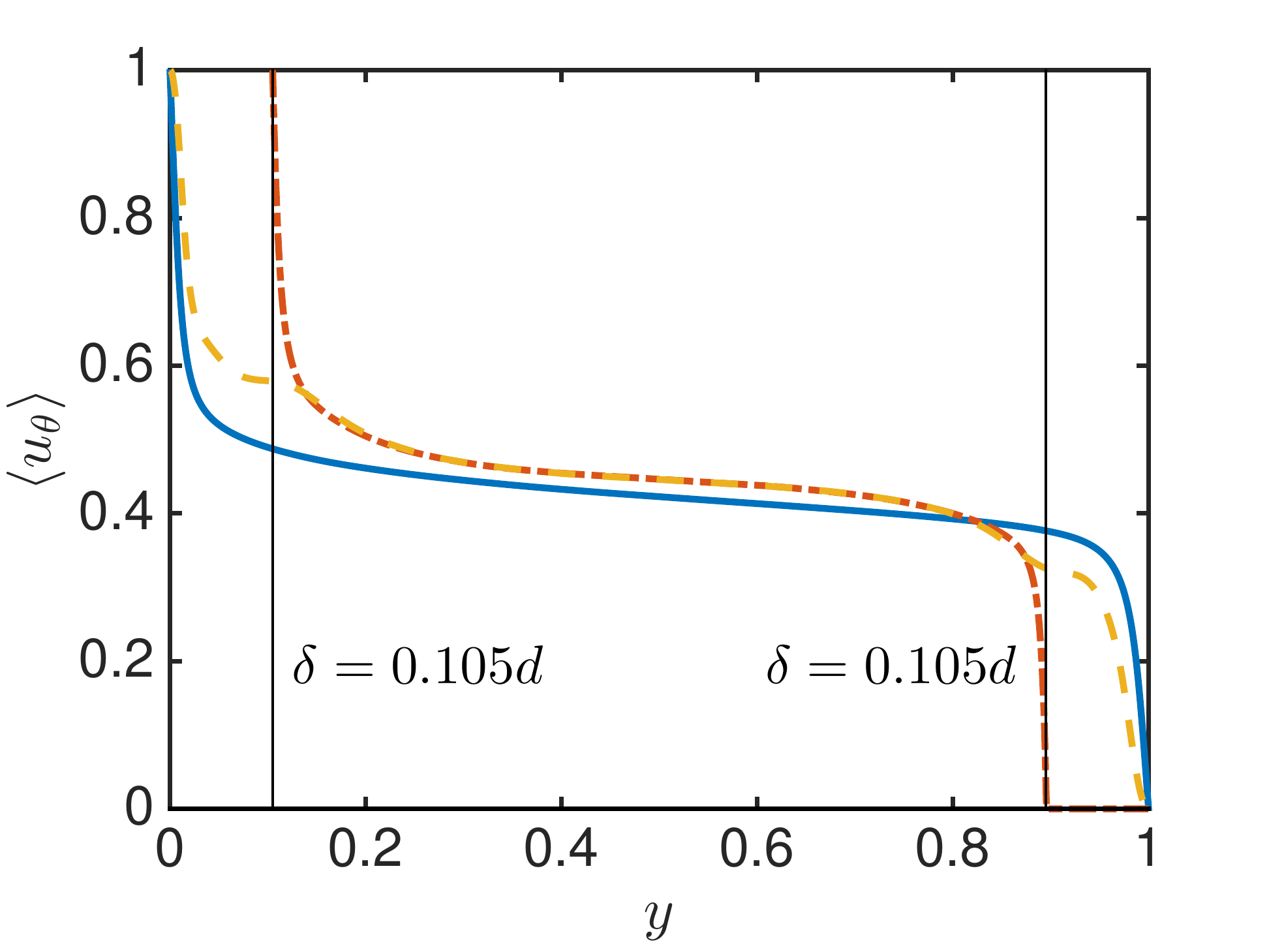}
    \centering {\par (b) \par}
  \end{minipage}
\caption {Averaged azimuthal velocity profile along radius for the smooth and the $\delta=0.105d$ case at two different Taylor numbers: (a) $Ta=4.77\times10^7$; (b) $Ta=9.75\times10^8$. The solid line shows time-, azimuthally, and axially averaged velocity profile for the smooth case. The dashed line denotes the grooved case profile over time, azimuth, and ten valleys while the dot-dashed line time, azimuth, and ten tips, respectively. }
\label{mean}
\end{figure}

\begin{figure}
  \centering
  \begin{minipage}[c]{0.48\textwidth}
    \includegraphics[width=2.5in]{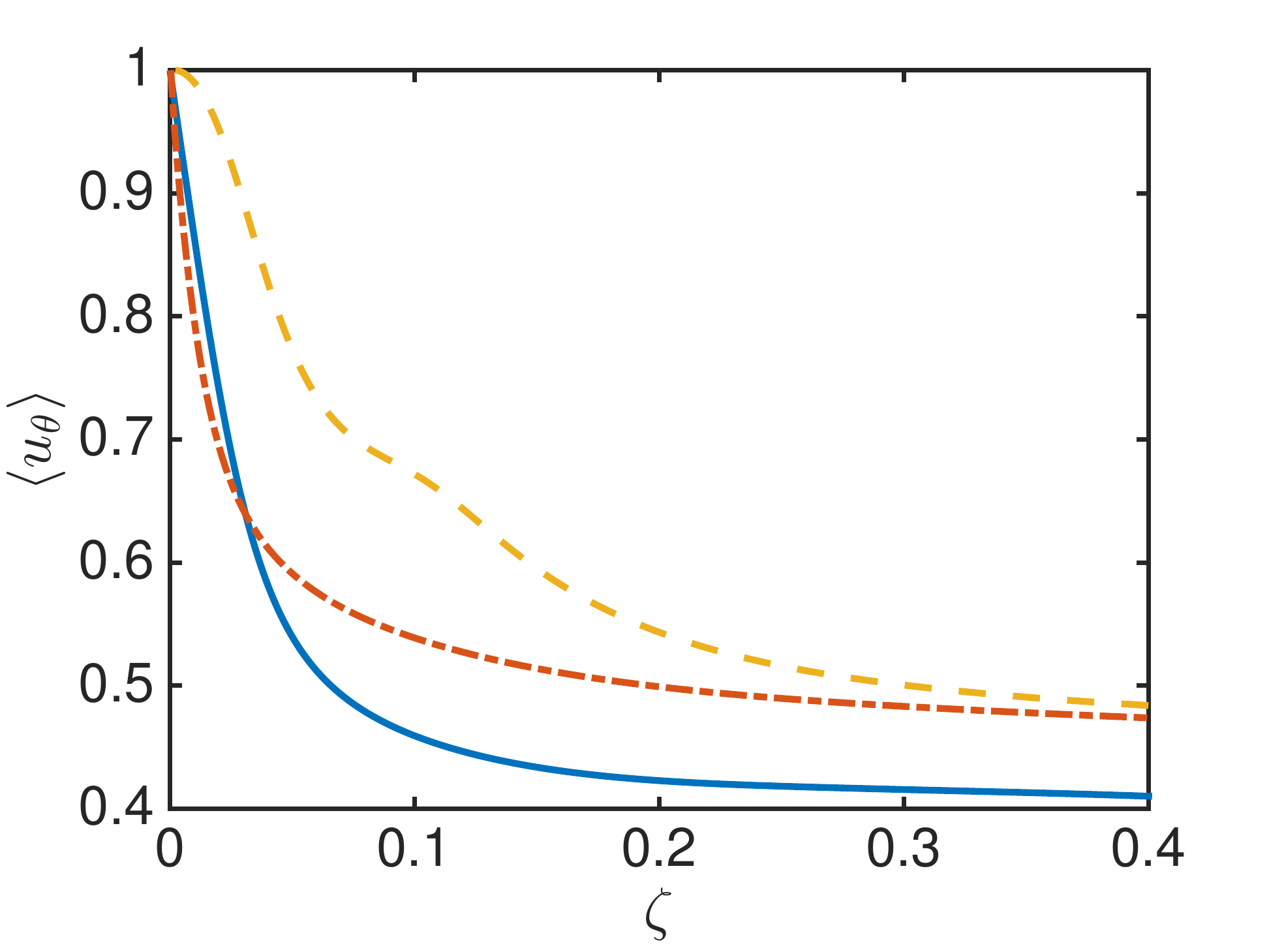}
    \centering {\par (a) \par}
  \end{minipage}%
     \begin{minipage}[c]{0.48\textwidth}
    \includegraphics[width=2.5in]{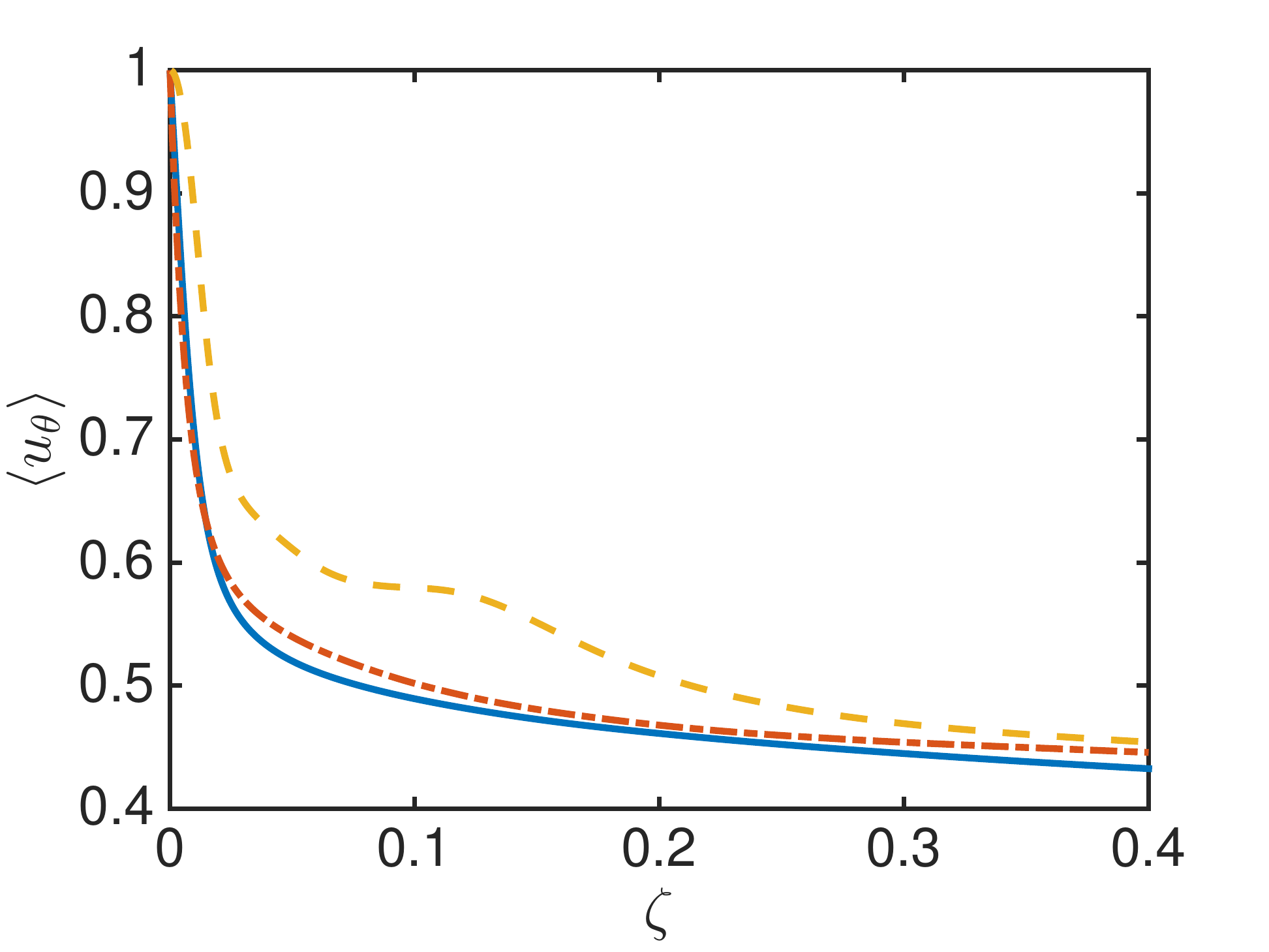}
    \centering {\par (b) \par}
  \end{minipage}
\caption {A close look at the near inner cylinder region for averaged azimuthal velocity profile of the smooth and the $\delta=0.105d$ case at two different Taylor numbers: (a) $Ta=4.77\times10^7$; (b) $Ta=9.75\times10^8$. The solid line shows time-, azimuthally, and axially averaged velocity profile for the smooth case. The dashed line denotes the grooved case profile over time, azimuth, and ten valleys while the dot-dashed line time, azimuth, and ten tips, respectively. Note that $\zeta=y-\delta/d$ is the horizontal distance from the solid surface.}
\label{meanBL}
\end{figure}

In this section we foucus the question of how exactly the grooves modify the BL dynamics. Figure \ref{mean} shows the mean velocity profiles in the smooth and in the $\delta=0.105d$ configuration at $Ta=4.77\times10^7$ and at $Ta=9.75\times10^8$. Around the smaller $Ta$ the scaling slope is larger than the effective ultimate regime scaling 0.38 while around the larger $Ta$ the effective scaling saturates back to 0.38. By choosing these two $Ta$, we can thus directly compare the BL differences between these two regimes. It is seen from this figure that the bulk velocity increases with the presence of wall roughness, i.e. the influence of the wall roughness penetrates well into the bulk region of the flow, because of the conservation of the angular velocity current $J^\omega=r^3(\langle u_r\omega \rangle_{A,t} - \nu \partial_r \langle \omega \rangle_{A,t})$ along the radius. Figure \ref{meanBL} shows an enlarged region of figure \ref{mean} near the wall of the inner cylinder. The velocity profile for the smooth case is time-, azimuthally, and axially averaged while for the grooved case it is time- and azimuthally averaged for ten different tip points and valley points of the grooves. For the smooth case, the velocity profile indicates that, due to turbulent mixing induced by the Taylor rolls, the azimuthal velocity is uniform in the bulk region and the velocity gradient across the radius is concentrated in thin boundary layers. With increasing $Ta$, the BL becomes thinner and the velocity gradient becomes steeper as expected. Correspondingly, the azimuthal velocity profile in the bulk becomes flatter and the bulk is more extended towards the walls. 

Similar features also apply to TC flow with grooves but at the same time it also shows some interesting characteristic differences. For the grooved case at $Ta=4.77\times 10^7$, the velocity gradient is larger at the tip of the groove while it attains the minimum value at the valley. This is because the flow trapped inside the groove is viscosity dominated and the secondary vortex is very weak. In comparison, at $Ta=9.75\times10^8$, the secondary vortex inside the groove is strengthened significantly. The strong secondary vortex fully mixes the flow inside the grooves and even a flat small bulk region can be seen. At the same time the BL thickness at the valley is greatly decreased as a result of the extension of the secondary vortex. Because the secondary vortex must flow smoothly over the valley where it is singular, the BL is thicker in the valley as compared to at the tip. We expect the difference of the BL thickness between the tip and valley point becomes smaller and smaller as increasing the $Ta$. The asymmetry of the BL at inner and outer surface is caused by the curvature of the cylinders and also depends on the strength of the Taylor rolls, which has been detailed in \cite{ostilla2014b}.

 \begin{figure}
\centering
\includegraphics[width=2.5in]{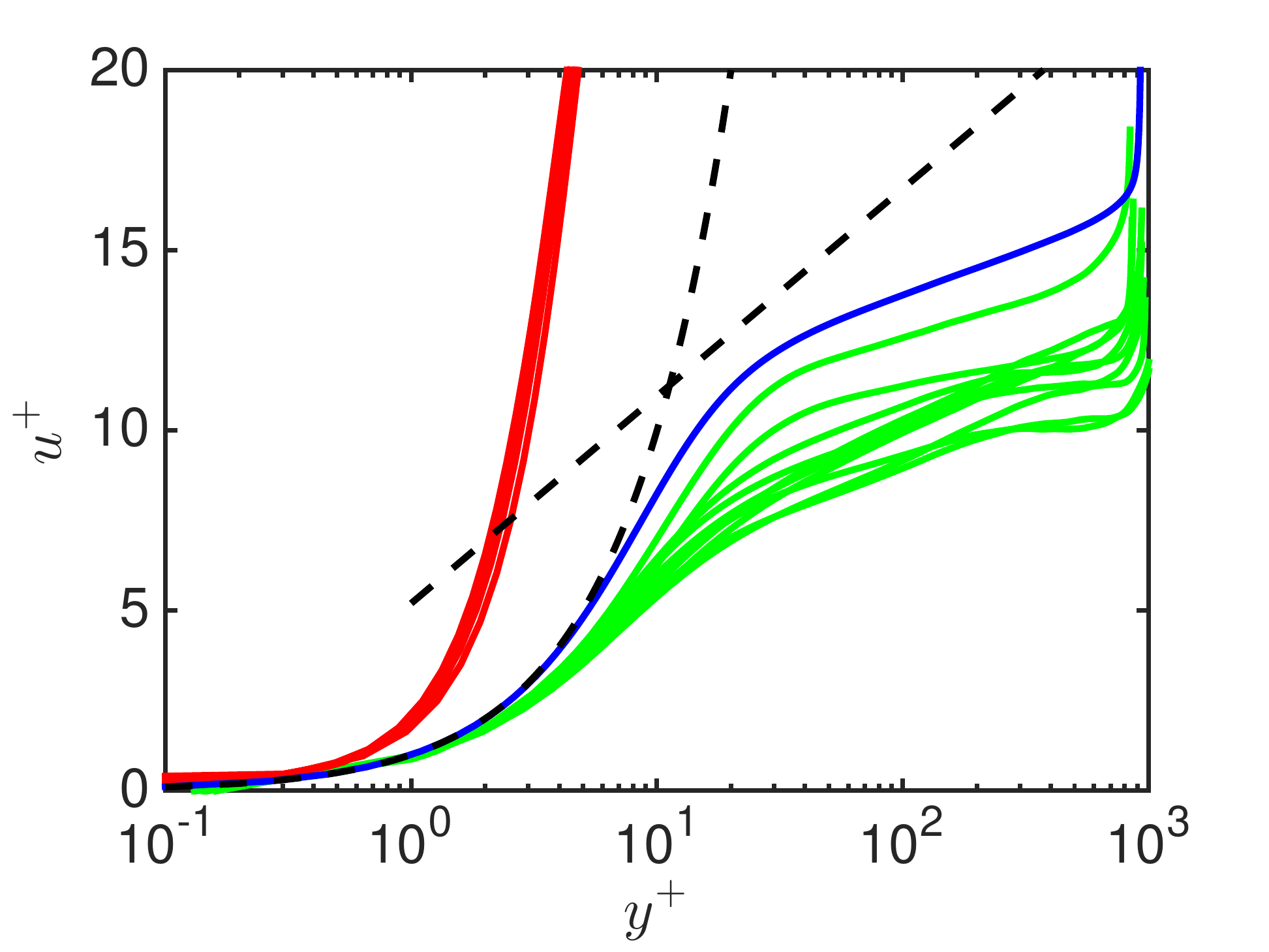}
    \caption{Velocity profiles of the smooth and the $\delta=0.105d$ case non-dimensionalized by friction velocity and wall distance at $Ta=9.75\times10^8$. The ten red and ten green lines represent the non-dimensionalized azimuthal velocity $u^{+}=(U-\langle u_\theta \rangle_{\theta,t})/u_*$ over ten valleys and ten tips for the grooved case, which show different interaction with the Taylor rolls. The local friction velocity is defined as $u_*=(\nu \langle \partial_r u_\theta(r_i) \rangle_{t,\theta})^{1/2}$, with $\partial_r$ the derivative normal to the
wall and $y^+$ is the non-dimensional wall distance $y^+=(r-r_i)u_*/\nu$ for valleys and $y^+=(r-r_i-\delta)u_*/\nu$ for tips in wall units. The blue line shows the averaged mean velocity profile for the smooth case at the same $Ta$. In the smooth case, we also average over axial direction. That is replacing $\langle ... \rangle_{\theta,t}$ by $\langle ... \rangle_{\theta,t,z}$ in the above definitions and the wall distance is the same as one used for the valleys. The dashed lines show the relationships $u^+=y^+$ and $u^+=2.5\ln(y^+)+5.2$. }
\label{profile}
\end{figure}

The distinctive feature for ultimate regime is the turbulent BL. For wall distances much larger than the internal length scale and much smaller than the outer length scale,
the mean velocity profile has a logarithmic dependence on the distance to the wall. It has been shown that in a lot of canonical flows such as pipe, channel and BL flow the logarithmic velocity profiles exist. We refer the reader to the reviews by \cite{marusic2010} and \cite{smits2011} for a detailed introduction. Figure \ref{profile} presents the non-dimensionalized azimuthal velocity $u^{+}=(U-\langle u_\theta \rangle_{t, \theta})/u_*$ on tips and valleys as a function of the wall distance $y^+$ for the inner cylinder boundary layer at $Ta=9.75\times10^8$ for the grooved case. We define $u_*$, the local friction velocity, with $u_*=(\nu \langle \partial_r u_\theta(r_i) \rangle_{t,\theta})^{1/2}$ for valleys and tips, with $\partial_r$ the derivative normal to the
wall and $y^+$ the non-dimensional wall distance $y^+=(r-r_i)u_*/\nu$ for valleys and $y^+=(r-r_i-\delta)u_*/\nu$ for tips in wall units. As a comparison, in figure \ref{profile} we also plot the averaged mean velocity profile for the smooth case at the same $Ta$. That is by replacing $\langle ... \rangle_{t, \theta}$ with $\langle ... \rangle_{t, \theta, z}$ in the above definitions and the wall distance is the same as one used for valleys. In the smooth case, the mean velocity profile is first linear in the viscous region and after a buffer region becomes logarithmic. For the grooved case, significant downward shifts in the log-law are obtained near the tips, whereas significant upward shifts are obtained near the valley. Because of the strong plume ejection, all tips of grooves show logarithmic behaviour in the boundary layer, in accordance with the findings by \cite{vanderpoel2015} and \cite{ostilla2014a} which show that the BL is turbulent at the place where plumes are ejected. The implication is that specific layout of grooves with tips could locally induce turbulent BLs at specific points. Meanwhile, because Taylor rolls still exist, the plumes have to follow the direction of the Taylor rolls, which results in the different velocity profile slopes for the logarithmic region at different tips (also see the right panel of figure \ref{time_averaged}). In the valley, as described before, the BL is thicker than at the tip, so that the flow is more viscosity dominated. It is interesting to note that the velocity profiles nearly overlap at the valleys.This indicates that different secondary vortices at different height are homogenous.

\begin{figure}
  \centering
  \begin{minipage}[c]{0.48\textwidth}
    \includegraphics[width=2.5in]{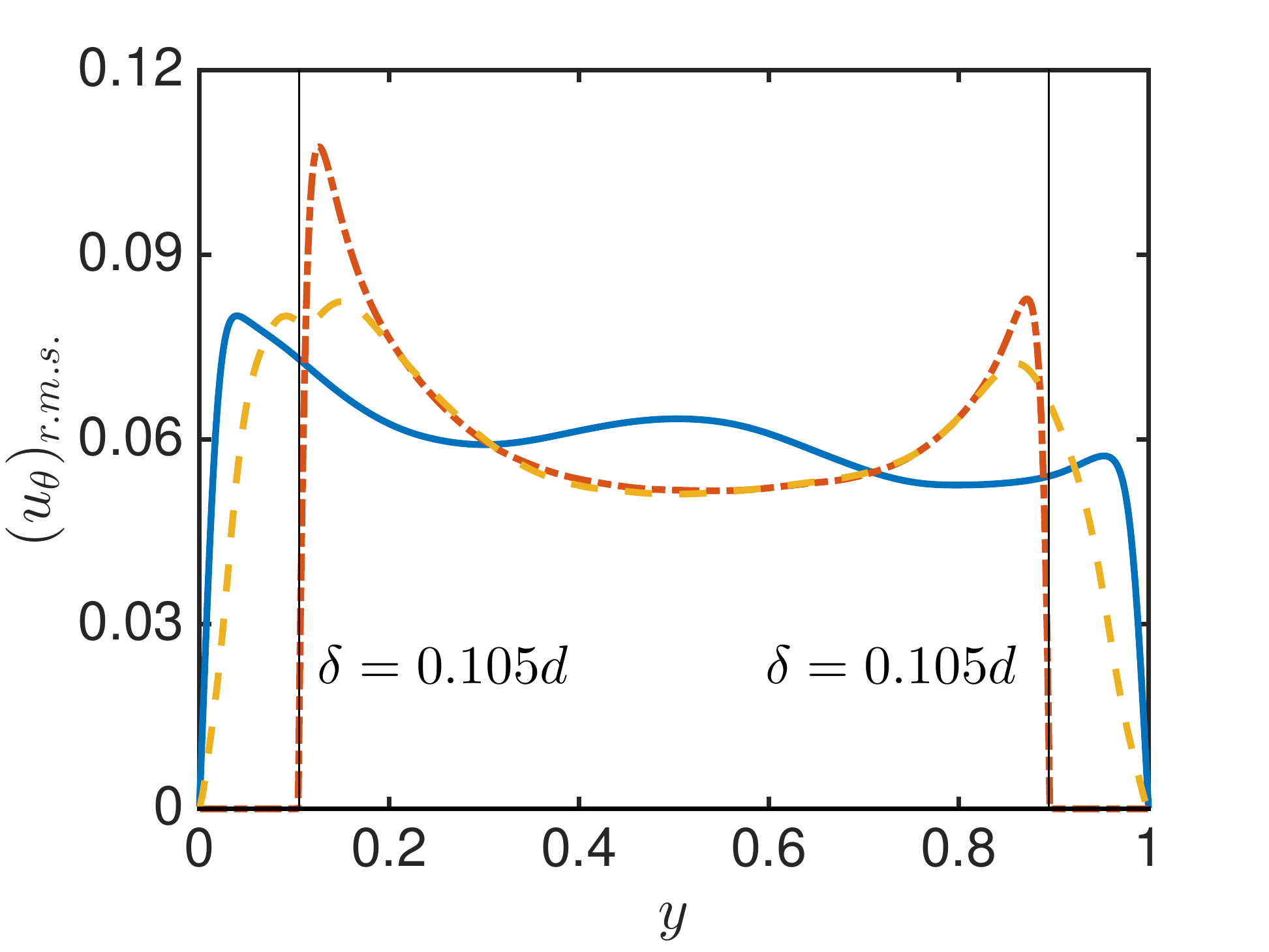}
    \centering {\par (a) \par}
  \end{minipage}%
     \begin{minipage}[c]{0.48\textwidth}
    \includegraphics[width=2.5in]{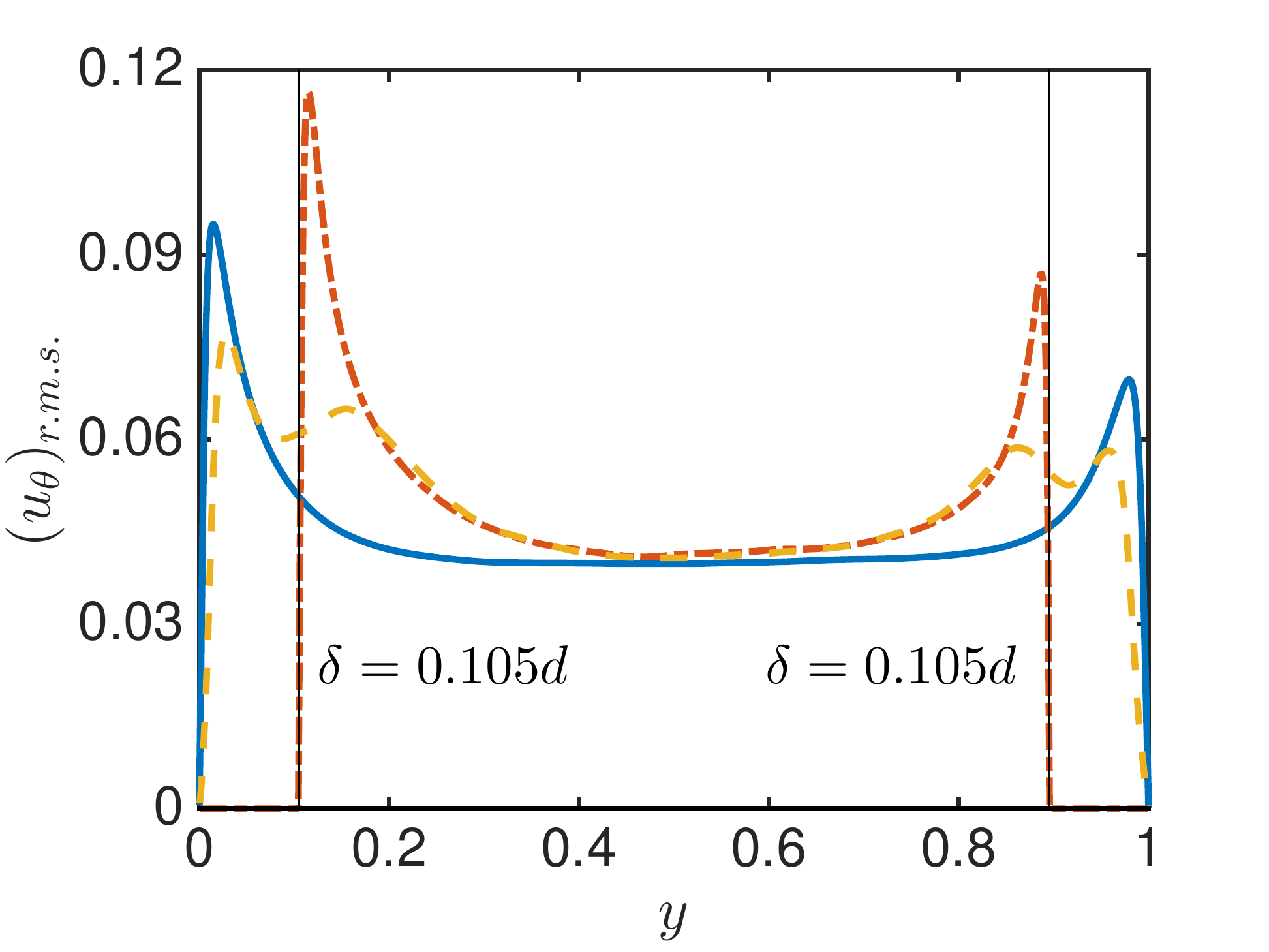}
    \centering {\par (b) \par}
  \end{minipage}
\caption {Root mean square azimuthal velocity profile of the smooth and the $\delta=0.105d$ case at two different Taylor numbers: (a) $Ta=4.77\times10^7$; (b) $Ta=9.75\times10^8$. The solid line shows time-, azimuthally, and axially averaged velocity profile for the smooth case. The dashed line denotes the grooved case profile for valleys while the dot-dashed denotes the profile above tips. The r.m.s velocity for the smooth case is defined as $(u_\theta)_{r.m.s}= (\langle u_\theta^2 \rangle _{t,\theta,z}- \langle u_\theta \rangle^2_{t,\theta,z})^{1/2}$ while for grooved case it is defined as $ (\langle u_\theta^2 \rangle _{t,\theta}- \langle u_\theta \rangle^2_{t,\theta})^{1/2}$ and then averaged over ten different tips or valleys. }
\label{RMS}
\end{figure}

\begin{figure}
  \centering
  \begin{minipage}[c]{0.48\textwidth}
    \includegraphics[width=2.5in]{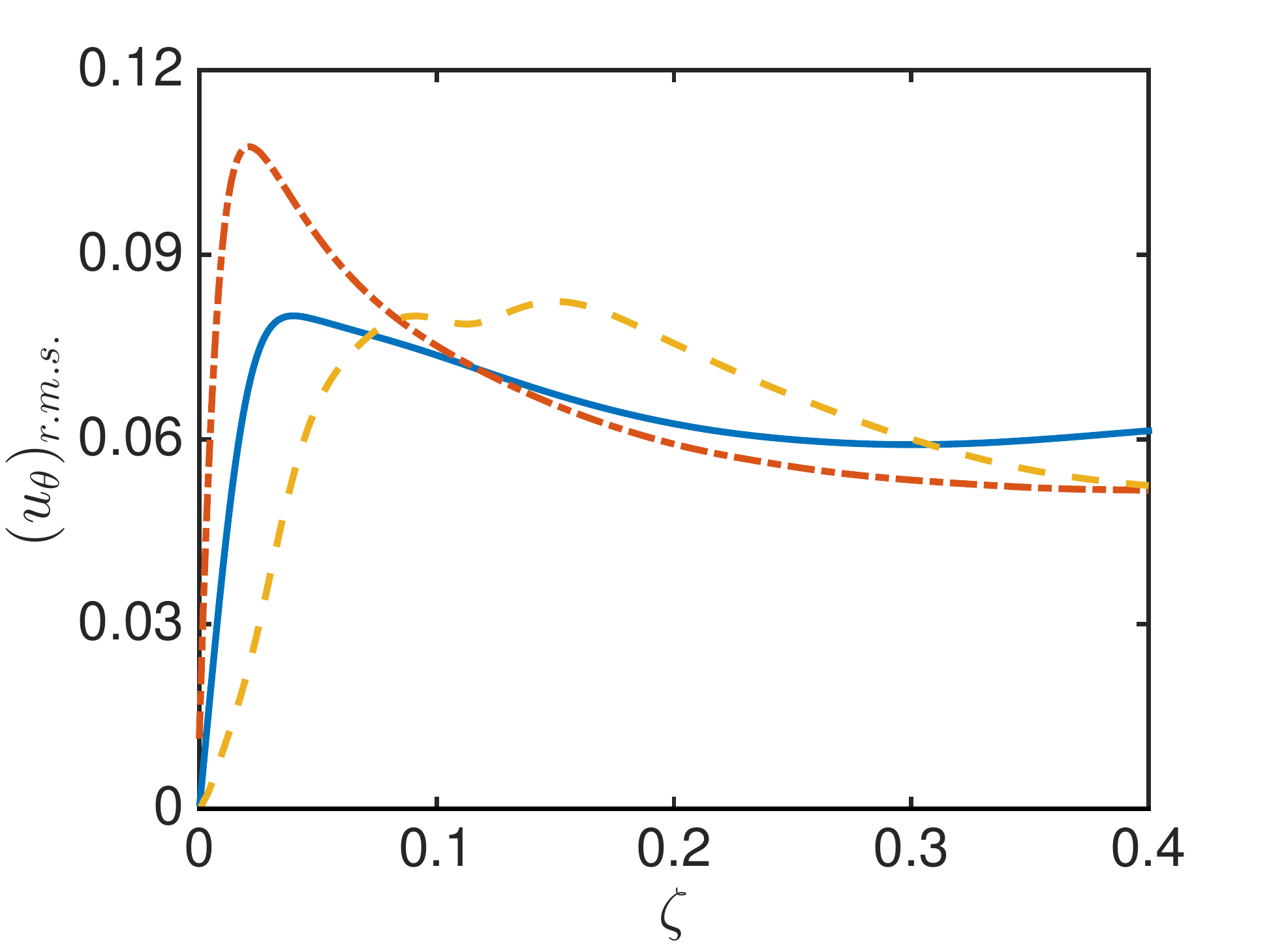}
    \centering {\par (a) \par}
  \end{minipage}%
     \begin{minipage}[c]{0.48\textwidth}
    \includegraphics[width=2.5in]{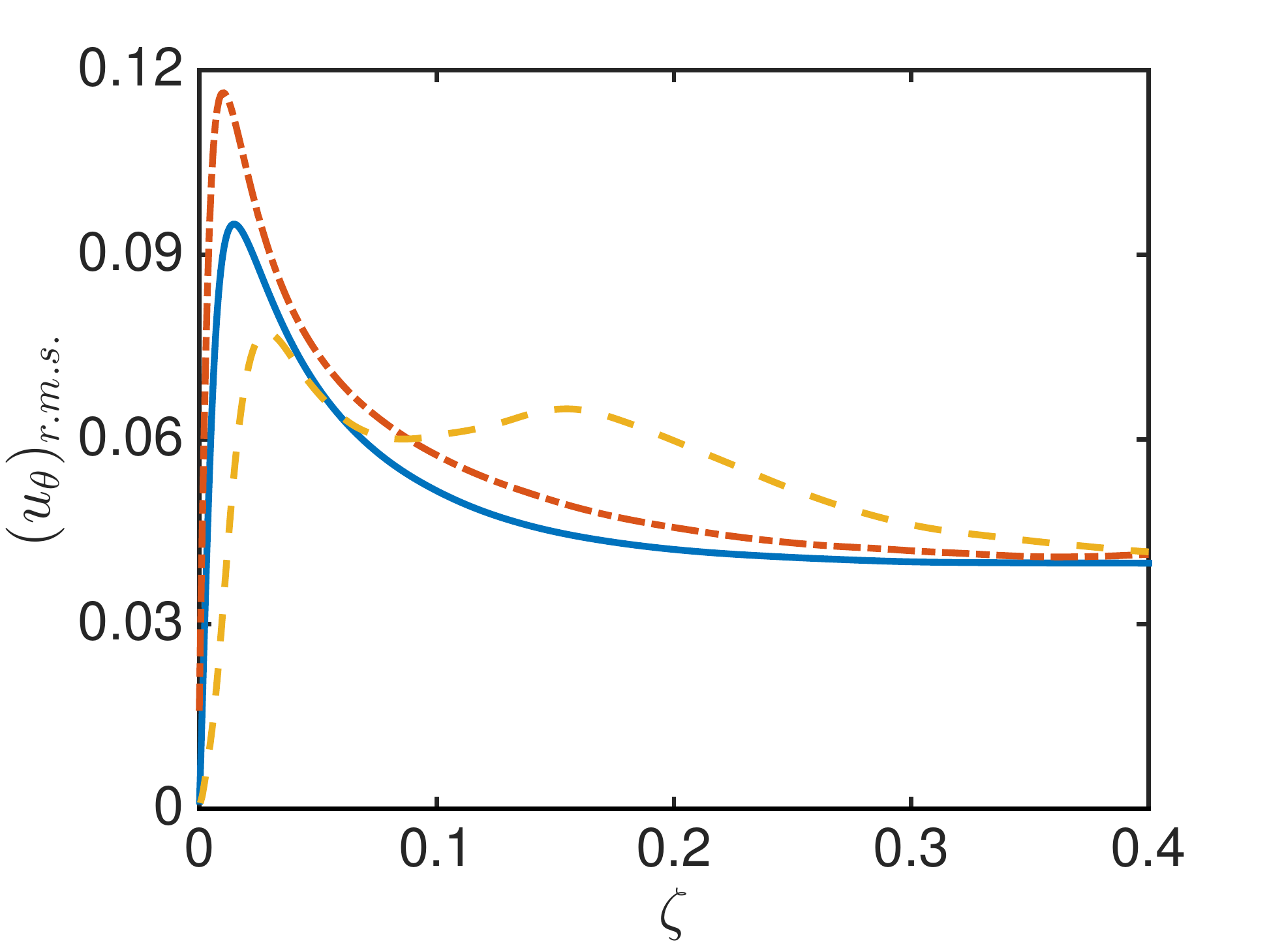}
    \centering {\par (b) \par}
  \end{minipage}
\caption {A close look at the root mean square azimuthal velocity profile of the smooth and the $\delta=0.105d$ case in the near wall region at two different Taylor numbers: (a) $Ta=4.77\times10^7$; (b) $Ta=9.75\times10^8$. The solid line shows time-, azimuthally, and axially averaged velocity profiles for the smooth case. The dashed line denotes the grooved case profile for valleys while the dot-dashed denotes the profile above the tips. The r.m.s velocity for the smooth case is defined as $(u_\theta)_{r.m.s}= (\langle u_\theta^2 \rangle _{t,\theta,z}- \langle u_\theta \rangle^2_{t,\theta,z})^{1/2}$ while for grooved case it is defined as $ (\langle u_\theta^2 \rangle _{t,\theta}- \langle u_\theta \rangle^2_{t,\theta})^{1/2}$ and then averaged over ten different tips or valleys. Note that $\zeta=y-\delta/d$ is the horizontal distance from the solid surface.}
\label{RMSBL}
\end{figure}

The grooves not only influence the mean velocity but also the fluctuations. Figure \ref{RMS} shows the root mean square azimuthal velocity profiles in the smooth and the $\delta=0.105d$ configuration at $Ta=4.77\times10^7$ and at $Ta=9.75\times10^8$. Figure \ref{RMSBL} shows an enlarged region of figure \ref{RMS} near the wall of the inner cylinder. The r.m.s velocity for the smooth case is defined as $(u_\theta)_{r.m.s}= (\langle u_\theta^2 \rangle _{t,\theta,z}- \langle u_\theta \rangle^2_{t,\theta,z})^{1/2}$ while for the grooved case it is defined as $ (\langle u_\theta^2 \rangle _{t,\theta}- \langle u_\theta \rangle^2_{t,\theta})^{1/2}$, averaged over different tips or valleys. For the smooth case, there is only one peak of the r.m.s velocity which is associated with the enhanced fluctuation in the buffer layer. With increasing $Ta$, as the turbulent intensity becomes larger and the BL becomes thinner as expected, the peak is shifted to the inner cylinder and the r.m.s is larger. 

\begin{figure}
  \centering
  \begin{minipage}[c]{0.48\textwidth}
    \includegraphics[width=3.0in]{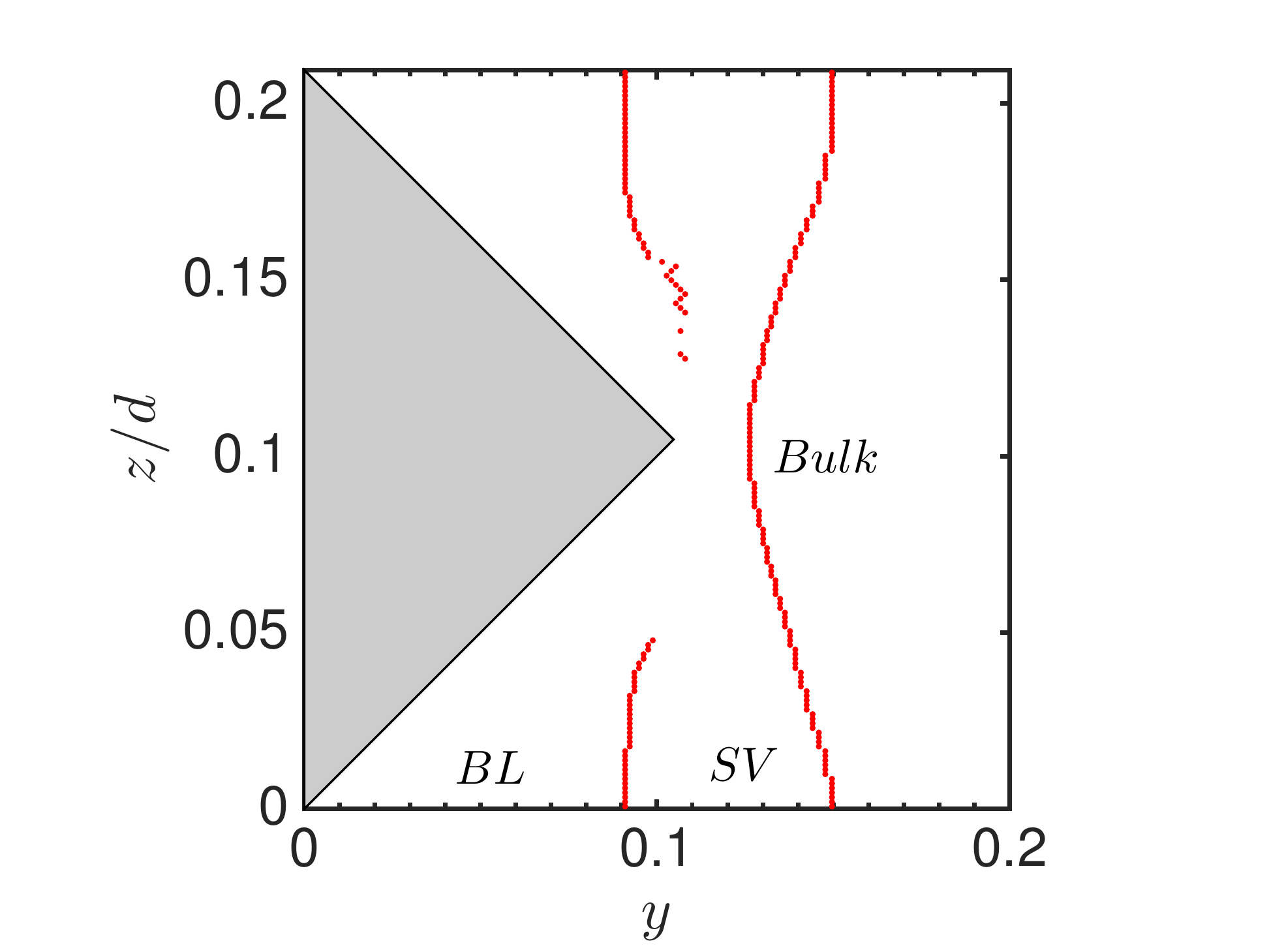}
    \centering {\par (a) \par}
  \end{minipage}%
     \begin{minipage}[c]{0.48\textwidth}
    \includegraphics[width=3.0in]{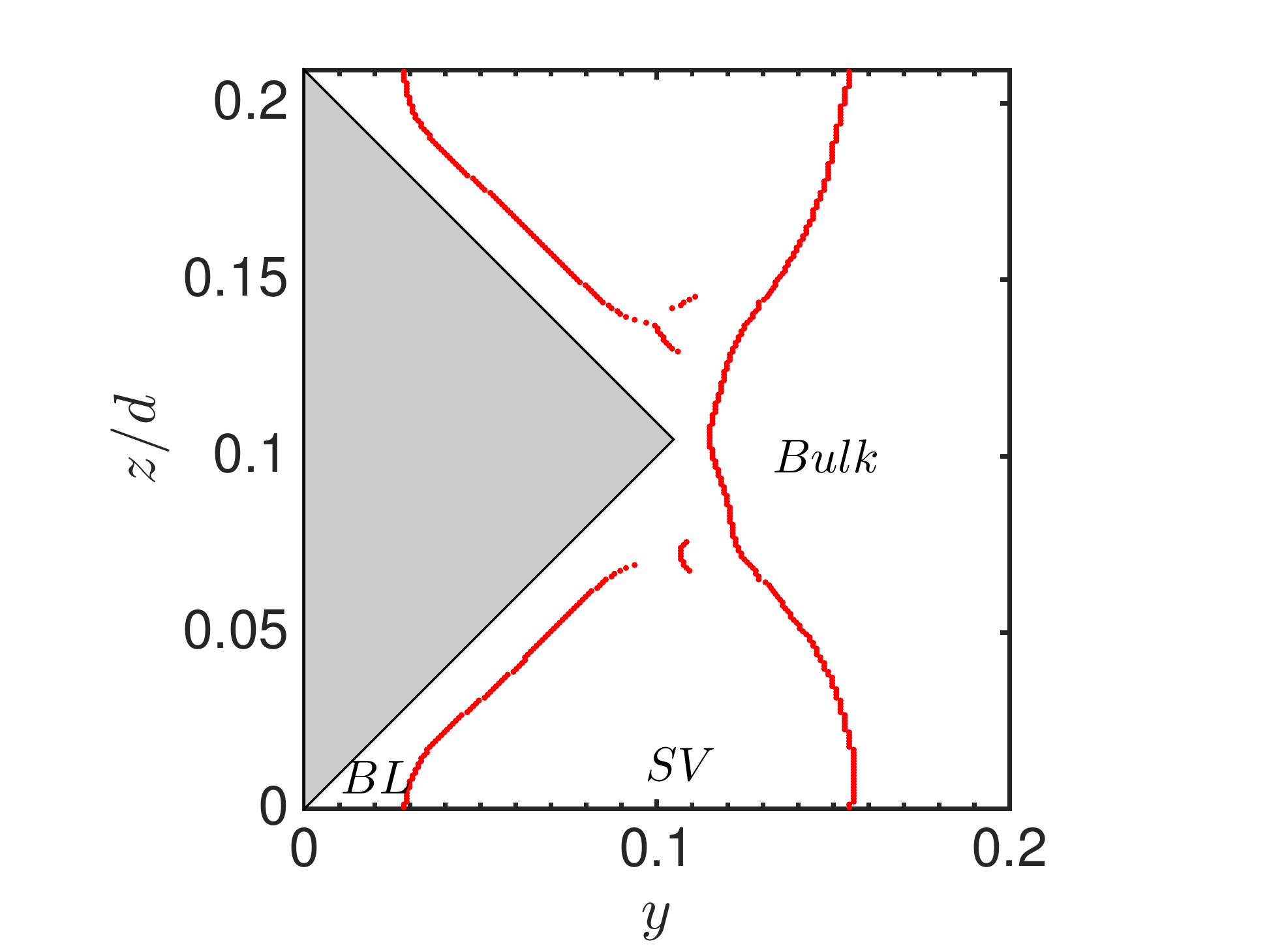}
    \centering {\par (b) \par}
  \end{minipage}
\caption {The positions of the peaks of the r.m.s azimuthal velocity profile near the inner cylinder for the $\delta=0.105d$ cases averaged in time, azimuth, and ten grooves at two different Taylor numbers: (a) $Ta=4.77\times10^7$; (b) $Ta=9.75\times10^8$. The red points are the positions of these peaks. We average over ten grooves. Abbreviations: BL$=$boundary layer, SV$=$secondary vortex.}
\label{BL}
\end{figure}

For the grooved case of $Ta=4.77\times10^7$, at tips of grooves, the position where the peak occurs is also shifted to the inner cylinder and the r.m.s velocity is larger. In the valley, we see two different peaks of the r.m.s. velocity. One close to the wall and one close to the bulk. The near wall one is associated with the buffer layer and the other is associated with the shear layer between the secondary vortex and the Taylor roll. The secondary vortex is located between these two peaks. Because the flow inside the groove is viscosity dominated and laminar, the position of this peak is far away from the wall and the intensity is less than in the shear layer. At $Ta=9.75\times10^8$, the secondary vortex becomes stronger and extends closer towards the wall, as illustrated before. As a result, the BL becomes thinner and shift closer to the wall at the same time the intensity of fluctuation becomes higher than the shear layer. 

 \begin{figure}
\centering
\includegraphics[width=2.5in]{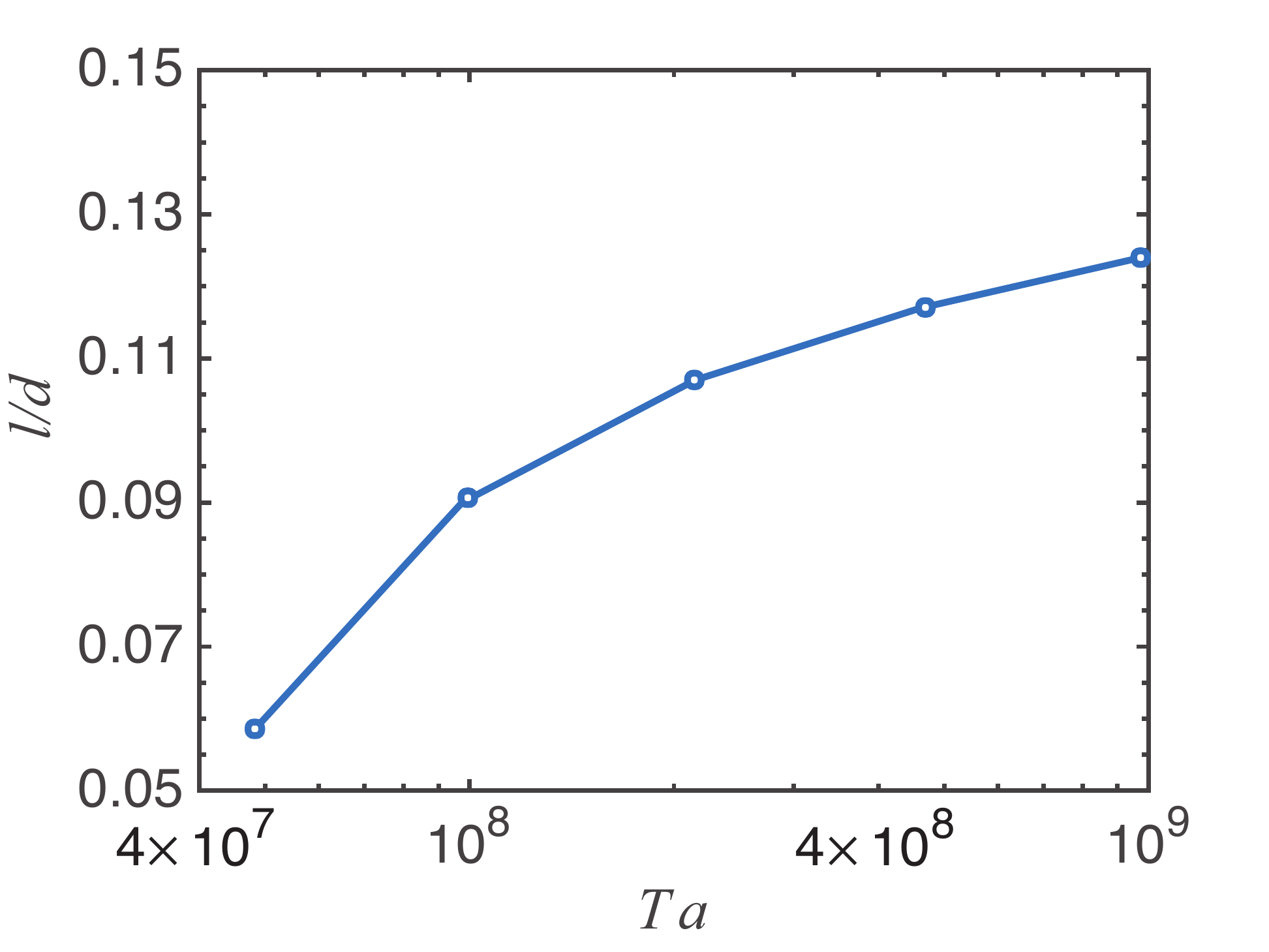}
    \caption{Distance $l/d$ between the peaks of the root mean square azimuthal velocity in the groove valley as a function of $Ta$ for the $\delta=0.105d$ cases.}
\label{sv}
\end{figure}

Figure \ref{BL} shows the positions where the peaks of the r.m.s azimuthal velocity profiles close to the inner cylinder occur for $Ta=4.77\times10^7$ and for $Ta=9.75\times10^8$, both for the $\delta=0.105d$ case. The profiles were averaged in the azimuthal direction, in time and over the ten grooves. Although other definitions are possible, inspired by the work of \cite{stringano2006}, we define the near wall peak of the r.m.s. velocity profile as the thickness of the boundary layer. This enables us to separate the flow domain into three zones. Between the wall and the near wall peak is the BL layer zone. Between the two peaks is the secondary vortex zone and beyond is the bulk zone. From the comparison between these two panels, it is seen that with increasing $Ta$, the BL inside the groove becomes thinner and more uniformly distributed along the surface of the groove. This indicates that at high $Ta$, the azimuthal velocity can not feel the effect of the grooves and thus forms the uniformly thick BL just as the smooth case. As an indication of the growing strength of the secondary vortices, in figure \ref{sv} the distance between the peaks of the r.m.s azimuthal velocity in the groove valley as a function of $Ta$ for $\delta=0.105d$ case is shown. It is found that the size of the secondary vortex saturates to 0.11d.

\begin{figure}
  \centering
  \begin{minipage}[c]{0.48\textwidth}
    \includegraphics[width=2.5in]{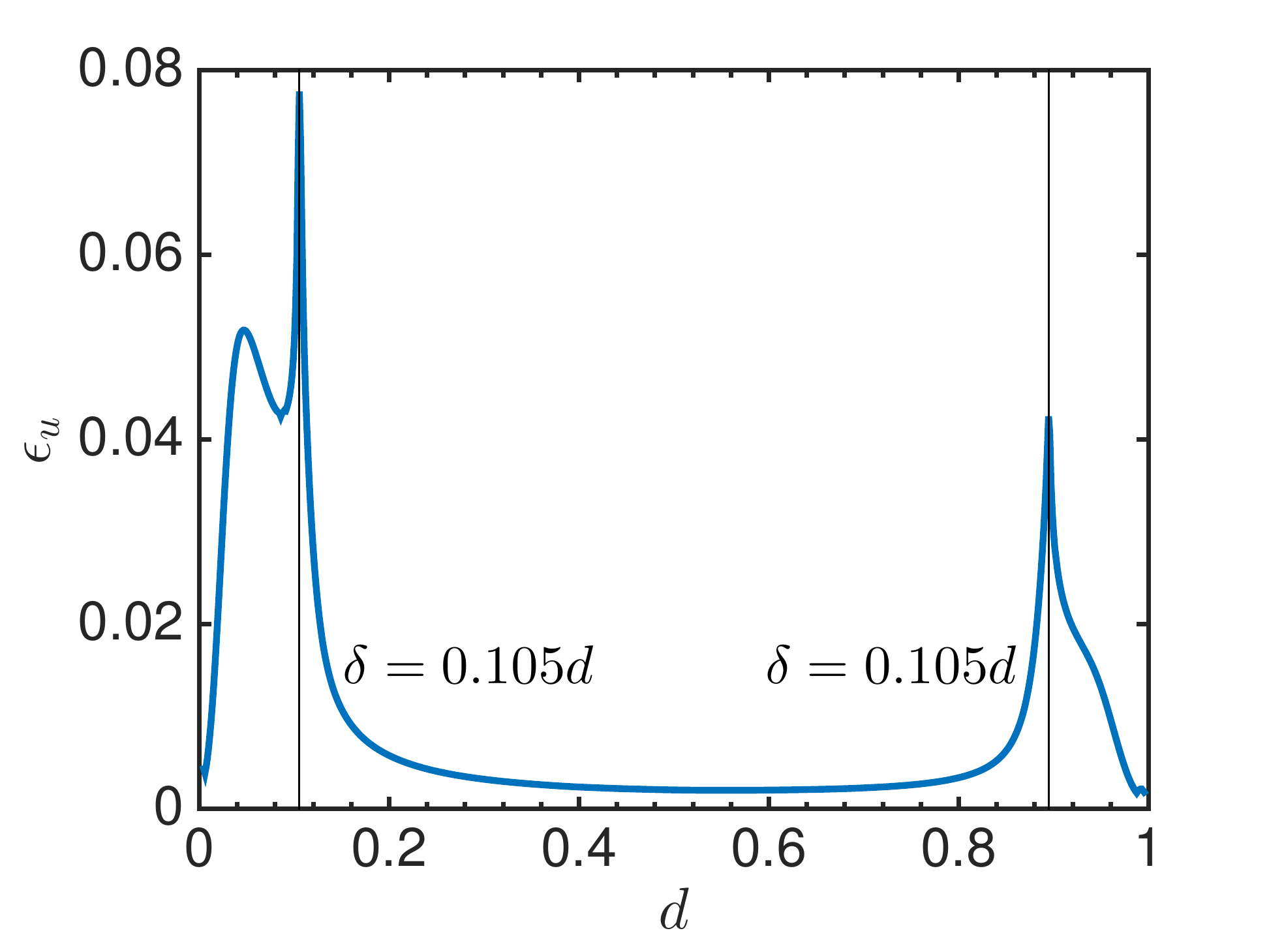}
    \centering {\par (a) \par}
  \end{minipage}%
     \begin{minipage}[c]{0.48\textwidth}
    \includegraphics[width=2.5in]{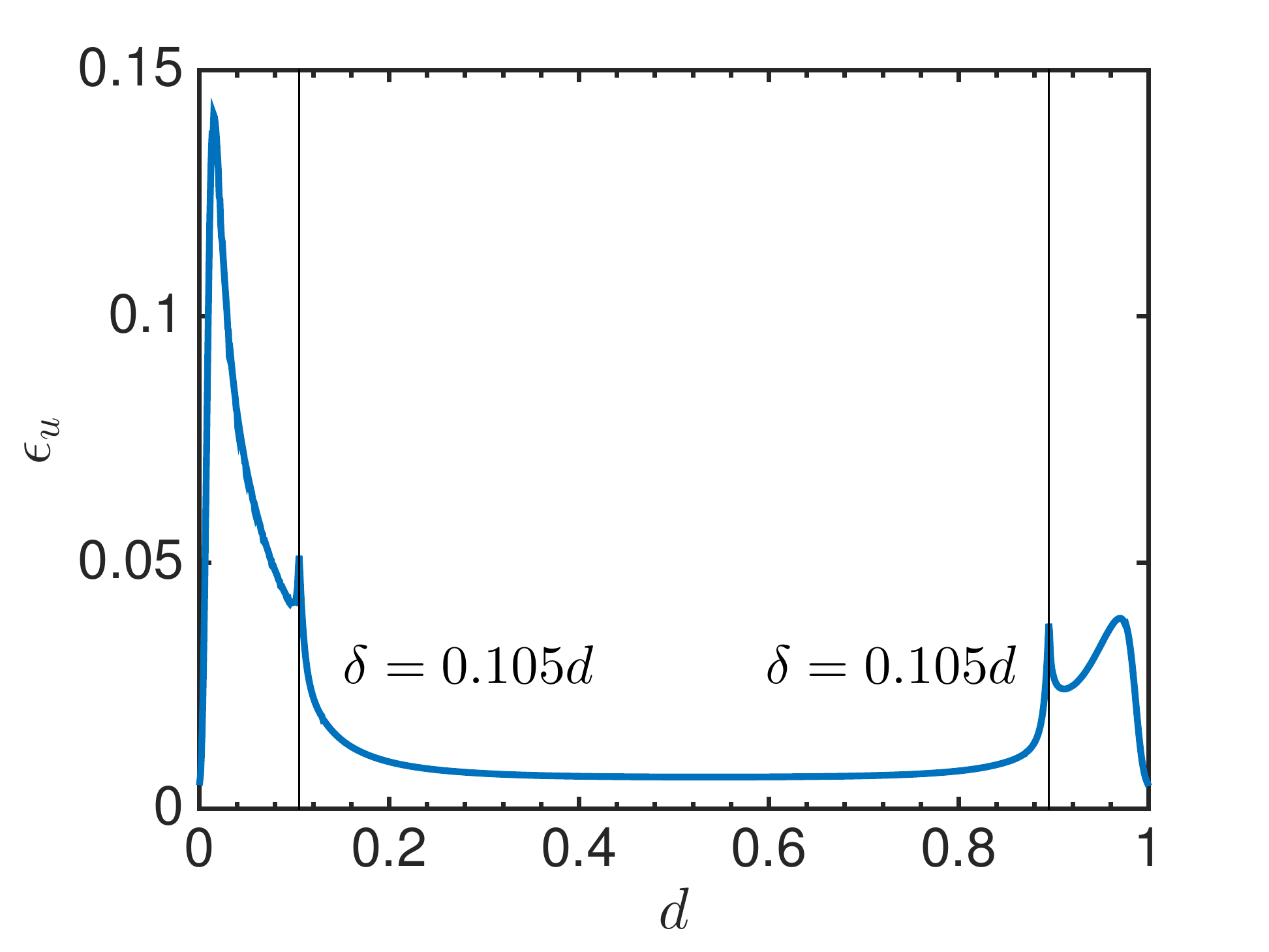}
    \centering {\par (b) \par}
  \end{minipage}
\caption {Energy dissipation rate $\epsilon_u=\nu \langle (\nabla {\bf{u}})^2  \rangle_{t, \theta, z}$ along the radius for the $\delta=0.105d$ series. It was averaged in time, azimuth, and height at two different $Ta$: (a) $Ta=4.77\times10^7$; (b) $Ta=9.75\times10^8$.}
\label{dissipation}
\end{figure}

Finally, to shed further light on why the effective scaling is larger than 0.38 at $Ta=4.77\times10^7$ and saturates back to 0.38 at $Ta=9.75\times10^8$ for the $\delta=0.105d$ series, we compare the energy dissipation rate $\epsilon_u=\nu \langle (\nabla {\bf{u}})^2  \rangle_{t, \theta, z}$ along the radius as shown in figure \ref{dissipation}. According to \cite{grossmann2011}, the logarithmic correction, which originates from the turbulent boundary layer, leads to an effective scaling of 0.38. Even in the fully turbulent regime in which we are, the boundary layer effects still play a major role on the scaling exponent. Only asymptotically, the $Nu$ vs. $Ta$ scaling exponent will go to 1/2, but we are far away from this exponent in any experiment. At $Ta=9.75\times10^8$, because the BL inside the groove turns to be more similar to the smooth case, the contribution of the BL to the energy dissipation returns. As a result, the logarithmic correction for the scaling law occurs. At $Ta=4.77\times10^7$, the BL inside the groove is thick and laminar, however, this is not the smooth laminar BL where there is a steep shear layer very close to the wall which contributes very much to the dissipation. In this case, the shear is very small in the valley region but large around the tip (see figure \ref{shear} and figure \ref{mean}). It is clearly seen that the shear rate in the region between the grooves is much less than that of the smooth case. Thus it is reasonable that the dissipation from the BL contribute less and it is bulk dominant. We estimate the dissipation contributed from BL and bulk at the two Taylor number from figure \ref{dissipation}. At $Ta = 4.77\times10^7$, the region below the groove height consists 49\% of the total dissipation while the bulk contributes 51\%. In contrast, at $Ta = 9.75\times10^8$, the region below the groove height consists 65\% of the total dissipation while the bulk contributes 35\%. Therefore, according to the GL theory \citep{grossmann2000}, in the regime around $Ta=4.77\times10^7$ where it is more bulk dominant, the local effective scaling exponent is larger than 0.38.

\section{Conclusions}
\label{sec6}

In this study, direct numerical simulations are conducted to explore the Taylor-Couette flow under the presence of grooved walls. Numerical results corresponding to the Taylor number up to $Ta=2.15\times10^9$ are presented for three different sizes of grooves, namely $\delta=0.052d$, $\delta=0.105d$, and $\delta=0.209d$, at a radius ratio $\eta=0.714$.

We find three different characteristic regimes reflecting in different effective scaling laws between $Nu_\omega$ and $Ta$. First, when $Ta<Ta_{th}$, i.e. when the boundary layer thickness is larger than the height of the grooves, there is a overlap regime in which smooth and groove cases show the same behaviour. Second, when $Ta>Ta_{th}$, i.e. when the boundary layer thickness is less than the groove height, there is a steep slope regime in which the power law exponent between $Nu_\omega$ and $Ta$ becomes larger than the ultimate region effective scaling 0.38. Third, when $Ta$ is large enough, there is a saturation regime in which the effective scaling law saturates back to 0.38. It is found that even after saturation the slope is the same as for the smooth case, the absolute value of torque is increased beyond the ratios of the surface area increase between grooved and smooth wall.


The visualization of the flow structure shows that the enhanced transport is caused by the plume ejection from the tips of the grooves. Firstly, the axial flow induces the secondary vortices inside the grooves. Secondly, the interaction between the secondary vortices and the Taylor vortices facilitates the flow separation on the tips. Thirdly, this flow separation causes the boundary layer to detach into the bulk, and thus the boundary layer flow forms the plumes and follows the preferential direction of Taylor vortices. Finally, the combination of plumes and flow separation greatly strengthens the convective part of $Nu_\omega$. In particular, there is the possibility that the torque can become smaller when there are no plumes ejected from the tips of the grooves and the grooves impede the Taylor rolls. Another interesting feature revealed from visualizations is that large scale Taylor vortices still survive under the presence of the grooves. This is because the induced secondary vortices inside the grooves also favour the circulation of the Taylor rolls. 

With increasing $Ta$, the intensity of the secondary vortex inside the grooves is strengthened. The boundary layer thickness in the valley is decreased and more uniformly distributed along the surface of the groove. A small flat bulk region for the mean velocity profile can be seen inside the grooves. At high $Ta$, the azimuthal velocity cannot feel the effect of the grooves and thus forms the uniformly thick BL just as the smooth case.
As to the fluctuation, it is found that on tips of grooves, there is only one peak for the root mean square azimuthal velocity while in the valley there are two peaks. The first is associated with the BL near the valley and the second with the shear layer between the secondary vortices and the Taylor vortices. We found that the steep slope regime is more bulk dominant and therefore the effective scaling slope is larger while in the saturation regime, the boundary layer contribution reoccurs and hence the scaling slope saturates.

Our ambition is to further understand plume triggered transitions. \cite{ostilla2014a} and \cite{vanderpoel2015} have shown that in smooth TC and RB the transition to ultimate regime can be triggered by plumes because the regions of BLs where the plumes are ejected become turbulent. On the one hand, we would like to study that whether in the grooved case the turbulent boundary layer can not only be formed on the tips of grooves but for much larger $Ta$ also inside the grooves and thus lead to the ultimate regime. On the other hand, grooves can be used to manipulate the plumes because they are ejected from the tips of the grooves. If we implemented more and more grooves and made the tips sharp enough so that there were more and more tips where plumes could be ejected, the ultimate regime possibly could be achieved at a much smaller $Ta$.

\bigskip
\textit{Acknowledgement}: This work is sponsored by the Foundation for Fundamental Research on Matter (FOM), which is a part of the Dutch organisation for Scientific Research (NWO). The authors would like to thank C. Sun, E. P. van der Poel and Y. Yang for many valuable discussions and the Dutch Supercomputing Consortium SurfSARA for the allocation of computing time. 

\bibliographystyle{jfm}
\bibliography{jfm-instructions}

\end{document}